\documentclass[a4paper,12pt]{article}
\usepackage{jheppub,esint,shuffle,psfrag}
 \usepackage[utf8]{inputenc}

\usepackage{esint} 
\usepackage{breqn}
\usepackage{bm}

\def \tr {\mathop{\rm tr}\nolimits}
\def \Im {\mathop{\rm Im}\nolimits}

\newcommand\lr[1]{{\left({#1}\right)}}

\newcommand \vev [1] {\langle{#1}\rangle}

\newcommand \ket [1] {|{#1}\rangle}
\newcommand \bra [1] {\langle {#1}|}
\newcommand\re[1]{(\ref{#1})}
\def \qqquad {\qquad\quad}
\def \qqqquad {\qquad\qquad}

\def\numberbysection{\@addtoreset{equation}{section}
                     \def\theequation{\thesection.\arabic{equation}}}

\begin{document}

\vspace{-4cm }

\begin{flushleft}
 \hfill \parbox[c]{30mm}{
 IPhT--T24/016}
\end{flushleft}
\author{Zoltan Bajnok$^a$, Bercel Boldis$^{a,b}$ and Gregory P. Korchemsky$^{c}$}
\affiliation{
$\null$
$^a$HUN-REN Wigner Research Centre for Physics, Konkoly-Thege Mikl\'os \'ut 29-33, 1121 Budapest, Hungary
\\
$\null$ 
$^b$
Budapest University of Technology and Economics  
 M\H{u}egyetem rkp. 3., 1111 Budapest, Hungary
\\		
$\null$
$^c${Institut de Physique Th\'eorique\footnote{Unit\'e Mixte de Recherche 3681 du CNRS}, Universit\'e Paris Saclay, CNRS,  91191 Gif-sur-Yvette, France}   
}
\title{
Solving four-dimensional superconformal Yang-Mills theories with Tracy-Widom distribution
}
\abstract{%
\small
We study a special class of observables in $\mathcal N=2$ and $\mathcal N=4$ superconformal Yang-Mills theories which, for an arbitrary 't Hooft coupling constant $\lambda$, admit representation as determinants of certain semi-infinite matrices. Similar determinants have previously appeared in the study of level-spacing distributions in random matrices and are closely related to the celebrated Tracy-Widom distribution. We exploit this relationship to develop an efficient method for computing the observables in superconformal Yang-Mills theories at both weak and strong coupling. The weak coupling expansion has a finite radius of convergence. The strong coupling expansion involves the sum of the `perturbative' part, given by series in $1/\sqrt\lambda$, and the `non-perturbative' part, given by an infinite sum of exponentially small terms, each accompanied by a series in $1/\sqrt\lambda$ with factorially growing coefficients.  We explicitly compute the expansion coefficients of these series and show that they are uniquely determined by the large order behavior of the expansion coefficients of the perturbative part via resurgence relations.}

\maketitle

%\tableofcontents
 
\section{Introduction}

In this paper we address the problem of computing observables in strongly coupled four-dimensional superconformal gauge theories. In the planar limit, 
these observables are given by nontrivial functions of the 't Hooft coupling constant $\lambda=g_{\rm YM}^2 N$. Determining them for an arbitrary $\lambda$ is an extremely difficult problem, and the number of known solutions is scarce. 

One such solution, particularly relevant for our analysis, involves calculating the expectation value of the half-BPS Wilson loop in maximally supersymmetric $\mathcal N=4$ super Yang-Mills theory (SYM).  By employing the localization technique~\cite{Pestun_2017}, it can be found  for arbitrary value of 't Hooft coupling constant and rank of the gauge group.
In the planar limit, the result is expressed in terms of a Bessel function
\begin{align}\label{W-exp}
\vev{W} = {2\over \sqrt\lambda} I_1(\sqrt\lambda) +O(1/N^2) \,.
\end{align}
This relation holds in planar $\mathcal N=4$ SYM for arbitrary $\lambda$. At weak coupling, $\vev{W}$ is given by a series in $\lambda$. Unlike generic Yang-Mills theories, where the weak coupling expansion has zero radius of convergence, this series 
has an infinite radius of convergence. This behavior is a consequence of the conformal symmetry of $\mathcal N=4$ SYM, which prevents the appearance of infrared renormalons, and the exponential suppression of instantons in the planar limit.

At strong coupling, the situation is different. Using the well-known properties of the Bessel functions one finds from \re{W-exp} that the asymptotic expansion of $\vev{W}$ takes the following form at large $g=\sqrt\lambda/(4\pi)$ 
\begin{align}\label{W-as}
\vev{W} = (4\pi g)^{-3/2} \left[  e^{4\pi g} \sum_{n\ge 0} {w_n\over g^n}  - i e^{-4\pi g}\sum_{n\ge 0} (-1)^n {w_n\over g^n}\right].
\end{align}
In contrast to the weak coupling expansion, the series on the right-hand side exhibit a zero radius of convergence. The expansion coefficients grow factorially at large orders $w_n\sim n!/(8\pi)^n$ and, as a consequence, the first series inside the brackets in \re{W-as} suffers from Borel ambiguities. Their contribution to the first term in \re{W-as} is exponentially small and scales as $O(e^{-4\pi g})$.   The Borel singularities are cancelled in the sum of the two terms, thus ensuring that $\vev{W}$ is well-defined for any 't Hooft coupling.  

The properties of $\vev{W}$ can be elegantly explained within the framework of the AdS/CFT correspondence \cite{Aharony:1999ti}. The strong coupling expansion \re{W-as} can be interpreted as a semiclassical expansion of the string partition function in the limit of vanishing string tension $\alpha'\sim 1/g$. The first term in \re{W-as} originates from a semiclassical calculation around a stable saddle point, representing the minimal area surface attached to a circle at the boundary of AdS. The second term in \re{W-as} arises from an unstable saddle point, hence the factor of $i=\sqrt{-1}$ in front of this term \cite{Drukker:2006ga}. It can be interpreted as a nonperturbative correction to the semiclassical expansion. While this correction is exponentially suppressed at strong coupling, it plays a crucial role in bridging the gap between the weak and strong coupling regimes.

The asymptotic expansion \re{W-as} was obtained from the known exact expression \re{W-exp}. Deriving this expansion directly from the AdS/CFT correspondence poses a number of technical challenges, and to date, it has only been possible to obtain the first few terms in \re{W-as} (see \cite{Medina-Rincon:2018wjs} and references therein).
Computing exponentially small corrections in $\mathcal N=4$ SYM at strong coupling is important for understanding the full dynamics of the theory. These corrections arise from non-perturbative effects that are not captured by traditional perturbative methods. A powerful tool for studying $\mathcal N=4$ SYM at strong coupling is integrability \cite{Beisert:2010jr}. It allows us to map the theory to an integrable quantum mechanical system, which can be solved exactly. By exploiting the integrability of the theory, precise expressions for the exponentially small corrections can be derived for various observables \cite{Alday:2007mf,Basso:2007wd,Basso:2009gh}.

In this paper, following \cite{Bajnok:2024epf}, we describe a general method for the systematic computation of a class of important observables in strongly coupled four-dimensional $\mathcal N=2$ and $\mathcal N=4$ superconformal gauge theories. The Wilson loop \re{W-as} is an illustrative example of this class of observables. Other observables involve
\begin{itemize}
\item The flux tube correlations in planar $\mathcal N=4$ SYM theory~\cite{Beisert:2006ez,Belitsky:2019fan,Basso:2020xts};
\item The four-point correlation function of infinitely-heavy half-BPS operators in the same theory \cite{Coronado:2018ypq,Coronado:2018cxj,Kostov:2019stn,Bargheer:2019kxb,Kostov:2019auq,Belitsky:2019fan,Bargheer:2019exp,Belitsky:2020qrm,Belitsky:2020qir};
\item  The leading nonplanar corrections to the free energy on the sphere and the circular Wilson loop  in $\mathcal N=2$ SYM \cite{Beccaria:2020hgy,Beccaria:2021vuc,Beccaria:2023kbl}.
\end{itemize}
A distinguished feature of these observables is that for arbitrary 't Hooft coupling they can be expressed as determinants of certain semi-infinite matrices
\begin{align}\label{F(g)}
e^{\mathcal F(g)} = \det\left(\delta_{nm}-K_{nm}(g)\right)\bigg|_{1\le n,m <\infty}\,,
\end{align}
where $g=\sqrt\lambda/(4\pi)$. The explicit form of the interaction matrix $K_{nm}(g)$ depends on both the chosen gauge theory and the observable of interest. It is specified below in Section~\ref{sect2}. 

The relation \re{F(g)} holds for arbitrary coupling constant $g$. In a free theory, for $g=0$, the matrix elements $K_{nm}(0)$ vanish leading to $\mathcal F(0)=0$. At weak coupling, the function $\mathcal F(g)$ can be expanded  over traces of powers of the interaction matrix 
\begin{align}\label{F-weak}
\mathcal F(g) = -\tr K(g) -\frac12 \tr K^2(g) -\frac13 \tr K^3(g) + \dots
\end{align}
The terms in the expansion \re{F-weak} become progressively less important because they are proportional to higher powers of the coupling constant.
Consequently, for a given order in $g^2$, the function $\mathcal F(g)$ can be approximated by the first few terms in \re{F-weak}. We show below that the weak coupling expansion of $\mathcal F(g)$ has a nonzero radius of convergence. Its value depends on the choice of observable. 

At finite coupling, all the terms in  \re{F-weak} become equally important and cannot be ignored. We describe below a method that allows us to systematically take them into account and compute $\mathcal F(g)$ for arbitrary coupling. We demonstrate that $\mathcal F(g)$ admits an asymptotic expansion similar to that of a logarithm of the Wilson loop \re{W-as}. In particular, the function $\mathcal F(g)$ is given by a transseries
\begin{align}\label{ill}
\mathcal F(g) = \mathcal F_0(g) + e^{-8\pi g x_1} \mathcal F_1(g)+ e^{-8\pi g x_2} \mathcal F_2(g)+\dots \,,
\end{align}
where the values of the exponents $0< x_1<x_2<\dots $ depend on the choice of observable. For instance, in the special case of the circular Wilson loop \re{W-as}, the corresponding function $\mathcal F_{\text{W}}(g) = \log \vev{W}$ takes the form \re{ill} with $x_n=n$. The first term on the right-hand side of \re{ill} represents the semiclassical approximation. 
The exponentially small terms in \re{ill} can be interpreted as nonperturbative corrections to $\mathcal F(g)$.

The relation \re{ill} involve the coefficient functions $\mathcal F_n(g)$ (with $n\ge 0$). They are given by series in $1/g$ ~\footnote{More precisely, 
for some observables the expansion starts at order $O(g)$ and 
the leading function $\mathcal F_0(g)$  contains logarithmically enhanced $O(\log g)$ terms, see \re{F-gen} below.}
\begin{align}
\mathcal F_n(g)=\sum_{k} \mathcal F^{(k)}_{n} \, g^{-k}\,,
\end{align}
where the expansion coefficients grow factorially at high orders, $\mathcal F^{(k)}_{n}\sim k!$. As a consequence, these functions suffer from Borel ambiguities making individual terms in \re{ill} ill-defined. For the function $\mathcal F(g)$ to be well-defined, the coefficient functions in \re{ill}
have to satisfy nontrivial resurgence relations \cite{Marino:2012zq,Dorigoni:2014hea,Aniceto:2018bis}. We compute the functions $\mathcal F_n(g)$ for various observables in
$\mathcal N=2$ and $\mathcal N=4$ superconformal Yang-Mills theories mentioned above and 
demonstrate that they indeed verify these relations.
 
The paper is organized as follows. In Section~\ref{sect2} we present the explicit expressions for the interaction matrix in \re{F(g)} corresponding to the various observables mentioned above.  
The weak coupling expansion of \re{F-weak} is discussed in Section~\ref{sect3}. We show that each term on the right-hand side of \re{F-weak} has a logarithmic cut that starts at $g^2=-1/16$, or equivalently $\lambda=-\pi^2$. In Section~\ref{sect4} we describe the technique to derive the strong coupling expansion of the function $\mathcal F(g)$ defined in \re{F(g)}. 
We apply it in Section~\ref{sect5} to compute the coefficient functions in \re{ill}. In Section~\ref{sect6}, we derive the explicit expressions for these functions for various observables. In Section~\ref{sect7}, we demonstrate that the coefficient functions in \re{ill} satisfy the necessary resurgence relations, ensuring that the function $\mathcal F(g)$ is well-defined. Some details of the calculations are presented in Appendix \ref{app}.

\section{Interaction matrix}\label{sect2}

The observables under consideration are given by determinants of a semi-infinite matrix \re{F(g)}.
The interaction matrix $K_{nm}(g)$ entering \re{F(g)} depends on the coupling constant 
\begin{align}
g={\sqrt\lambda\over 4\pi}\,.
\end{align}
Surprisingly, its matrix elements take a universal form across all observables described in the Introduction. They admit an integral representation
\begin{align} \label{eq:K_nm}
K_{nm}(g)=\int_{0}^{\infty}dx\,\psi_{n}(x)\chi\Big(\frac{\sqrt{x}}{2g}\Big)\psi_{m}(x)\,,
\end{align}
which involves two different functions, $\chi(x)$ and  $\psi_n(x)$ (with $n\ge 1$).  
The former function is conventionally called the symbol of the matrix. It is assumed to vanish for $x\to\infty$.~\footnote{More refined conditions for $\chi(x)$ are formulated below, see \re{chi-Phi}.} Note that the dependence of \re{eq:K_nm} on the coupling constant resides in the argument of $\chi ( {\sqrt{x}}/(2g) )$. This function suppresses the contribution to \re{eq:K_nm} from large $x\gg (2g)^2$ and plays the role of a cut-off.

The functions $\psi_n(x)$ are given by the normalized Bessel functions 
\begin{equation}\label{psi}
\psi_{n}(x)=(-1)^{n}\sqrt{2n+\ell-1}{J_{2n+\ell-1}(\sqrt{x})\over\sqrt{x}}\,,
\end{equation}
where the argument of the Bessel function and the normalization factor are chosen  to simplify their orthogonality condition
\begin{align}\label{ortho}
\int_{0}^{\infty}dx\,\psi_{n}(x)\psi_{m}(x)=\delta_{nm}\,,\qquad (n,m\ge 1)\,.
\end{align}
The functions $\psi_n(x)$ and, as a consequence, the function $\mathcal F(g)$ also depend on a non-negative integer $\ell$. Its value is specified below.

For $\chi(x)=1$ the matrix \re{eq:K_nm} simplifies as $K_{nm}(g)=\delta_{nm}$ and the determinant \re{F(g)} vanishes.  For $\chi(x)=\theta(1-x)$, it yields a nontrivial function of the coupling constant $\mathcal F(g)$. The same function is famously known to describe the eigenvalue spacing at the hard edge of the Laguerre unitary ensemble \cite{Forrester:1993vtx}. It coincides with the Tracy-Widom distribution \cite{Tracy:1993xj} and can be computed explicitly in terms of solutions to the Painlev\'e~V equation.

In general, the function $\mathcal F(g)$ defined in \re{F(g)} and \re{eq:K_nm} depends on both the symbol function $\chi(x)$ and the value of $\ell$. 
In the context of four-dimensional SYM theories, their specific form is dictated by the choice of the observable. We encounter four different cases of $\chi(x)$ and $\ell$
\begin{align}\notag\label{chis}
& \chi_{\text{W}}(x) = - {(2\pi)^2\over x^2}\,, && \hspace*{-10mm} \ell_{\text{W}}=2\,,
\\[2mm]\notag
& \chi_{\text{f.t.}}(x) = 1- \coth(x/2)\,,&&  \hspace*{-10mm} \ell_{\text{f.t.}}=0, 1\,,
\\[2mm]\notag
& \chi_{\text{oct}}(x|y,\xi) = {\cosh y + \cosh \xi\over \cosh y + \cosh \sqrt{x^2+\xi^2}}\,,&&  \hspace*{-10mm} \ell_{\text{oct}}=0\,,
\\ 
& \chi_{\text{loc}}(x) =- {1\over\sinh^{2}(x/2)} \,, &&  \hspace*{-10mm} \ell_{\text{loc}}=1, 2\,.
\end{align}
Here the subscript refers to the observable it describes. 

Specifically, the determinant \re{F(g)} evaluated for $\chi(x)$ and $\ell$ as defined on the first line in \re{chis} yields the expectation value of the circular Wilson loop \re{W-exp} in planar $\mathcal N=4$ SYM theory. For an arbitrary integer $\ell\ge 3$, it yields the expectation value of the product of the circular Wilson loop and half-BPS operator with scaling dimension $\Delta=\ell-1$ (see \re{F-exact} below). The second line in \re{chis} corresponds to the flux tube correlations~\cite{Beisert:2006ez,Belitsky:2019fan,Basso:2020xts}. The third line in \re{chis} corresponds to the octagon form factor \cite{Kostov:2019stn,Kostov:2019auq,Belitsky:2020qrm,Belitsky:2020qir}. Its square gives the simplest four-point correlation function of infinitely-heavy half-BPS operators \cite{Coronado:2018ypq,Coronado:2018cxj}.~\footnote{In general, this correlation function
admits a factorization into a product of two octagon form factors with an arbitrary bridge length $\ell$.} The latter depends on two cross ratios $y$ and $\xi$ which enter the definition of $\chi_{\rm oct}$ in \re{chis}.  The last line in \re{chis} corresponds to the calculation of the free energy on the sphere and the circular Wilson loop  in $\mathcal N=2$ SYM theory using the localization technique~\cite{Beccaria:2020hgy,Beccaria:2023kbl}.
 
\subsection*{Symbol function}

The four symbol functions in \re{chis} have some properties in common:
\begin{itemize}
\item The function $1-\chi(x)$ has a definite parity under $x\to -x$ and is either strictly positive or strictly negative for $x\ge 0$.
\item It also has a power-like behaviour at the origin
\begin{align}\label{chi-orig}
1-\chi(x) =  b\, x^{2\beta} \left[1+O(x^2)\right],
\end{align}
where the exponent $\beta$ takes integer or half-integer values depending on the symbol in \re{chis}
\begin{align}\label{betas}
\beta_{\text{W}} = -1\,,\qquad \beta_{\text{f.t.}} = -\frac12\,,\qquad \beta_{\text{oct}} =1\,,\qquad \beta_{\text{loc}} =-1\,.
\end{align}
\item For all but the first symbol in \re{chis}, the function $\chi(x)$ decreases exponentially fast at infinity,  
\begin{align}\label{chi-inf}
\chi(x)\stackrel{x\to\infty}{\sim} c\, e^{-x}\,.
\end{align}
\end{itemize}
The first symbol in \re{chis} instead, has power-like behaviour at infinity. As we show below, this difference has an interesting manifestation in the properties the function \re{F(g)} at weak coupling. We also observe that the last three symbol functions in \re{chis} are related to each other as
\begin{align}\label{chi-rel}
  1-\chi_{\text{loc}}(x) = {1\over 1-\chi_{\text{oct}}(x|0,0) }=\lr{1-\chi_{\text{f.t.}}(x)}^2\,,
\end{align}
where the octagon function is evaluated for $y=\xi=0$. 

Taking into account the above properties, we can treat all four functions in \re{chis} in a unified manner by considering the following Wiener-Hopf type
 parametrization of the symbol function
\begin{align}\label{chi-Phi}\notag
1-\chi(x) {}&= b \,x^{2\beta} \prod_{n\ge 1} {1+x^2/(2\pi x_n)^2\over 1+x^2/(2\pi y_n)^2}
\\
{}&= b \,x^{2\beta} \Phi(x) \Phi(-x)\,.
\end{align}
Here $\beta$ is (half) integer and $\Phi(x)$ is a meromorphic function  analytical in the upper half-plane $\Im x>0$. It takes the following general form
\begin{align}\label{Phi}
 \Phi(x) =\prod_{n\ge 1} {1-{ix\over 2\pi x_n}\over 1-{ix\over 2\pi y_n}} 
\end{align}
and depends on an infinite set of real positive parameters $x_n$ and $y_n$ (with $n\ge 1$). They define the position of zeros and poles of the function $ \Phi(x)$, located at $x=-2\pi ix_n$ and $x=-2\pi i y_n$, respectively. 
 
The symbol function \re{chi-Phi} is specified by positive parameters $x_n$ and $y_n$ (with $n\ge 1$).
To simplify the analysis, we can assume without loss of generality that both sets of parameters are ordered increasingly, $0< x_1 < x_2< \dots$ and  $0< y_1 < y_2< \dots$.
The condition for $\chi(x)$ to satisfy \re{chi-inf} leads to nontrivial relations for  the large $n$ behaviour of $x_n$ and $y_n$.  For $x_n\sim (n-\beta_-)$ and 
$y_n\sim (n-\beta_+)$ one finds from  \re{chi-inf} and \re{chi-Phi} that $\beta=\beta_+-\beta_-$.   

Combining together \re{chis} and \re{chi-Phi}, it is straightforward to identify the functions $\Phi(x)$ for each symbol in \re{chis}
\begin{align}\label{cases}\notag
{}& \Phi_{\text{W}}(x) = 1-{ix\over 2\pi}\,,
\\ 
{}& \Phi_{\rm loc}(x) ={1\over \Phi_{\rm oct}(x|0,0)} =[\Phi_{\text{f.t.}}(x)]^2 =  \pi \left[{\Gamma(1-{ix\over 2\pi})\over\Gamma({1\over 2}-{ix\over 2\pi})}\right]^2.
\end{align}
The relations between the functions $\Phi(x)$ on the second line follow from \re{chi-rel}. The relation for $\Phi_{\rm oct}$ only holds for $y=\xi=0$. For generic $y$ and $\xi$ the expression for $\Phi_{\rm oct}(x|y,\xi)$ is more complicated. To save space we do not present it here.

The functions \re{cases} have the expected form \re{Phi}.
The function $\Phi_{\text{W}}(x)$ can be obtained from \re{Phi} by setting $x_1=1$ and sending all remaining $x_{n}$ and $y_n$ to infinity.
As follows from \re{cases},  the functions $\Phi_{\rm loc}(x)$ and $\Phi_{\text{f.t.}}(x)$ share the common sets of poles and zeros. Moreover, the poles of $\Phi_{\rm loc}(x)$ coincide with the zeros of $\Phi_{\rm oct}(x|0,0)$, and vice versa,  
\begin{align}\notag\label{xs}
{}& x_n^{\rm f.t.}=x_n^{\rm loc}=y_n^{\rm oct}=n-\frac12\,,\qqqquad 
\\[2mm]
{}& y_n^{\rm f.t.}=y_n^{\rm loc}=x_n^{\rm oct}=n  \,,
\end{align}
where $n\ge 1$. Notice that the poles and zeros of $\Phi_{\rm loc}(x)$ and $\Phi_{\rm oct}(x|0,0)$ are double degenerate. This property plays an important role in what follows. 

\section{Weak coupling expansion}\label{sect3}

In this section, we derive the weak coupling expansion of the determinant \re{F(g)}. For $g\to 0$, the integration region in \re{eq:K_nm} effectively shrinks to a point and the matrix $K_{nm}(g)$ vanishes in this limit leading to
\begin{align}\label{F-w}
\mathcal F(g)=O\Big(g^{2(\ell+1)}\Big)\,.
\end{align}
Indeed, 
changing the integration variable in \re{eq:K_nm} as $x\to (2g)^2 x$, we find that the matrix elements scale as $K_{nm}(g)=O\lr{(g^2)^{n+m+\ell-1}}$. This allows us to expand \re{F(g)} in powers of the matrix and arrive at \re{F-weak}. The traces on the right-hand side of   \re{F-weak}  behave as $\tr(K^L) = O(g^{2L(\ell+1)})$ and, therefore, each subsequent term in the weak coupling expansion \re{F-weak} is suppressed relative to the preceding term by a factor of $g^{2(\ell+1)}$. 

The leading term in \re{F-weak} is given by~\cite{Belitsky:2020qir}
\begin{align}\label{trB}
\tr K(g) = g^{2(\ell+1)}\sum_{k=0}^\infty (-g^2)^{k}{q_{\ell+k+1} (2\ell+2k)!\over k! (2\ell+k)! (\ell+k+1)!^2} \,,
\end{align}
where the notation was introduced for
\begin{align}\label{qk}
q_k=2k \int_0^\infty dx\, x^{2k-1} \chi(x)\,.
\end{align}
The expansion \re{trB} is well-defined provided that the symbol function $\chi(x)$ decreases sufficiently rapidly at infinity, ensuring the convergence of the integral   \re{qk}. Taking into account \re{chi-inf} we find that $q_k$ are finite and grow factorially at large $k$  
\begin{align}\label{q-large}
q_k \sim c \int_0^\infty dx\, x^{2k}  e^{-x} = c \, (2k)!\,.
\end{align}
This relation holds true for all symbol functions in \re{chis} except the first one. The corresponding value of the normalization constant $c$ is  
\begin{align}\label{cc}
c_{\text{f.t.}}=-2\,,\qqqquad c_{\text{oct}}= 2(\cosh y+\cosh\xi)\,,\qqqquad c_{\text{loc}}=-4\,.
\end{align}
Note that $c_{\text{oct}}\ge 4$ for real $y$ and $\xi$.

The function $\chi_{\text{W}}(x)=-(2\pi)^2/x^2$ does not satisfy \re{chi-inf} and the integral \re{qk} is divergent. In this case, we find from \re{eq:K_nm} that $K_{nm}(g) \sim g^2$ and the relation \re{trB} gets replaced with  
\begin{align}\label{trK}
\tr K_{\text{W}}(g) = -(4\pi g)^2\int_0^\infty {dx \over x}\psi_n(x)\psi_m(x) = -{(2\pi g)^2\over \ell}\,.
\end{align} 
By calculating the remaining terms in \re{F-weak}, it is possible to perform a resummation of the weak coupling expansion to all orders in $g^2$, thereby obtaining a closed-form expression for the determinant \re{F(g)} in terms of the modified Bessel function~\cite{Beccaria:2023kbl}
\begin{align}\label{F-exact}
e^{\mathcal F_{\text{W}}(g)} = \Gamma(\ell) (2\pi g)^{1-\ell} \,I_{\ell-1}(4\pi g)\,.
\end{align}
For $\ell=2$, the expression on the right-hand side coincides with the expectation value \re{W-exp} of half-BPS Wilson loop in planar $\mathcal N=4$ SYM.  For arbitrary $\ell\ge 3$, the relation \re{F-exact} coincides with the expression for the correlator $\vev{W O_{\ell-1}(x)}$ of the Wilson loop with a properly normalized half-BPS operator of dimension $\Delta=\ell-1$ \cite{Semenoff:2001xp}. 

The relation \re{F-exact} holds true for any coupling constant. Notably, its weak coupling expansion exhibits an infinite radius of convergence. We can show that for the remaining symbol functions in \re{chis} satisfying \re{chi-inf} the weak coupling expansion of the determinant \re{F(g)} has a finite radius of convergence.

To demonstrate this, we replace $q_{\ell+k+1}$ in \re{trB} with its large order behaviour \re{q-large} to get
\begin{align}
\tr K(g) \sim (-1)^\ell {c\over 4\pi} \log(g^2-g_\star^2)\,,
\end{align}
where $g_\star^2=-1/16$ or equivalently $\lambda_\star=-\pi^2$. In planar $\mathcal N=4$ SYM, this specific value of the 't Hooft coupling holds particular significance because it coincides with a singularity of the dispersion relation of excitations within the integrability approach \cite{Beisert:2010jr}.

We can show (see Appendix \ref{app} for details) that the remaining terms in \re{F-weak} also have a logarithmic singularity at $g^2=g_\star^2$
\begin{align}\label{KL}
\tr K^L(g) {}& \sim (-1)^{L(1+\ell)-1}   \frac{(c/4)^L \Gamma (L)}{\pi\,\Gamma^2\big(\frac{L+1}{2}\big)} \log(g^2-g_\star^2)\,.
\end{align}
Substituting this relation into \re{F-weak} we can perform the summation over $L$ to get
\begin{align} 
 \mathcal F(g) \sim  {1\over 2\pi^2}\arcsin(c/2) \left(\arcsin(c/2)+(-1)^{1+\ell} \pi \right) \log(g^2-g_\star^2)\,.
\end{align}
We recall that the parameter $c$ governs the asymptotic behaviour of the symbol function \re{chi-inf} and is given by \re{cc}.  
We observe that the function $\arcsin(c/2)$ develops an imaginary part for $|c|>2$. 

As follows from \re{cc}, this is the case for the last two symbol functions in \re{chis}. This indicates that the weak coupling expansion of the corresponding functions $\mathcal F_{\text{oct}}(g)$ and $\mathcal F_{\text{loc}}(g)$ possess additional singularities at values of $g^2$ that are closer to the origin than the previously identified value of $g_\star^2$.  Importantly, these singularities do not appear in any individual term on the right-hand side of \re{F-weak} but only arise when the expression is resummed.
      
\section{Strong coupling expansion}\label{sect4}   

At strong coupling, the interaction matrix \re{eq:K_nm} grows with $g$, rendering the expansion in \re{F-weak} ill-defined. In this section, we derive the strong coupling expansion of the determinant \re{F(g)} by exploiting  the relation between the interaction matrix \re{eq:K_nm} and the so-called truncated Bessel operator~\cite{BasorEhrhardt03}.

\subsection*{Semiclassical expansion}
 
As shown in \cite{Beccaria:2022ypy}, for the symbol function of the form \re{chi-Phi}, the strong coupling expansion of \re{F(g)} looks as
\begin{align}\label{F-gen}
{\cal F}(g)=-gA_{0}+\frac{1}{2}A_{1}^{2}\log g+B+f(g)+\Delta f(g)\,,
\end{align}
where the expansion coefficients depend both on the symbol function $\chi(x)$ and the nonnegative integer $\ell$. The leading asymptotic behaviour of ${\cal F}(g)$ is governed by the first three terms in \re{F-gen}.~\footnote{These terms can be interpreted as a generalization of the Szeg\H{o}-Akhiezer-Kac formula \cite{Bttcher2006AnalysisOT} for the semi-infinite Bessel matrix \re{eq:K_nm}.} The last two terms in \re{F-gen} vanish at large $g$. 

The first term on the right-hand side of \re{F-gen} grows linearly with $g$.  Its coefficient is given by the first Szeg\H{o} limit theorem \cite{Bttcher2006AnalysisOT}
\begin{align}\label{A00}
A_0=2\int_{0}^{\infty}\frac{dx}{\pi} x\partial_{x}\log(1-\chi(x))\,.
\end{align}
It is independent of $\ell$ and takes the following values  for the symbols defined in \re{chis}  
\begin{align}\label{A0}
A_0^{\text{W}} = -4\pi\,,\qqqquad
A_0^{\text{loc}} =-A_0^{\text{oct}}=2A_0^{\text{f.t.}}=-\pi\,.
\end{align}
Here $A_0^{\text{oct}}$ is evaluated for $\xi=y=0$. 
%Notice that the sign of $A_0$ follows the sign of $\chi(x)$: $A_0$ is negative when $\chi(x)<0$ and positive when $\chi(x)>0$.

The second term in \re{F-gen} grows logarithmically with $g$. Its origin can be attributed to the presence of a Fisher-Hartwig singularity in the symbol function \re{chi-Phi}.  The corresponding coefficient $A_1$ is given by
\begin{align}\label{A1}
A_1^2={\beta^2+2\ell\beta}\,.
\end{align}
It depends on the nonnegative integer $\ell$ and the parameter $\beta$ which defines the asymptotic behaviour of the symbol at the origin \re{chi-orig}. For the symbol functions in \re{chis} its value is given by \re{betas}.
 Notice that $A_1$ is universal in the sense that it does not depend on the choice of the $\Phi-$function in \re{chi-Phi}.

The third term in \re{F-gen} is independent of the coupling constant and is conventionally called the Widom-Dyson constant. This constant depends on the symbol function in a nontrivial way and its general expression can be found in \cite{Beccaria:2022ypy}. We  present below the explicit expressions for $B$ in the various cases in \re{chis}.

The forth term in \re{F-gen} involves the function $f(g)$ which is given by an asymptotic series 
\begin{equation}\label{f}
f(g)=\sum_{k=1}^{\infty}\frac{A_{k+1}}{2k(k+1)}g^{-k}\,.
\end{equation}
The expansion coefficients can be found to any order in $1/g$ using the method of differential equations \cite{doi:10.1142/S0217979290000504,Korepin:1993kvr,Tracy:1993xj}.
The first few coefficients are \cite{Belitsky:2020qrm,Belitsky:2020qir}
\begin{align}\notag\label{As}
 {}& A_{2}=-\frac{(4\ell_{\beta}^{2}-1)}{4}I_{1}\,, 
 \\\notag
 {}& A_{3}=-\frac{3(4\ell_{\beta}^{2}-1)}{16}I_{1}^{2}\,,
 \\
 {}&
A_{4}=-\frac{(4\ell_{\beta}^{2}-1)(16I_{1}^3+(4\ell_{\beta}^{2}-9)I_{2})}{128}\,,\quad \dots
 && 
\end{align}
where the notation was introduced for  $\ell_{\beta}=\ell+\beta$. 

In general, the coefficients $A_{k+1}$ in \re{f} are given by multi-linear combination of the functions $I_n=I_n(\chi)$ defined as
%~\footnote{The integral in \re{prof} is well-defined for the symbol function $\chi(x)$ satisfying \re{chi-orig} and decreasing sufficiently fast at infinity.} 
\begin{align}\label{prof}\notag
I_{n}(\chi) {}&=\int_{0}^{\infty}\frac{dx}{\pi}\frac{(x^{-1}\partial_{x})^{n}}{(2n-1)!!} x\partial_{x}\log(1-\chi(x))
\\
{}& =(-1)^{n-1}\sum_{k\ge 1} \left[{1\over (2\pi x_k)^{2n-1}}-{1\over (2\pi y_k)^{2n-1}}\right]\,.
\end{align}
Here in the second relation we replaced the function $\chi(x)$ with its general expression \re{chi-Phi}. 
For the symbol functions defined in \re{cases} the integrals \re{prof} can be expressed in terms of odd Riemann zeta values
\begin{align}\label{Is}\notag
{}& I_n^{\text{W}} ={(-1)^{n-1}\over (2\pi)^{2n-1}}\,,
\\\notag
{}& I^{\text{f.t.}}_{n=1}=\frac{\log 2}{2\pi }\,, \qqqquad    I_{n>1}^{\text{f.t.}} = (-1)^{n-1}(1-2^{2-2n}) {\zeta_{2n-1}\over \pi^{2n-1}} ,
\\[2mm]
{}& I_n^{\text{loc}} =-I_n^{\text{oct}}=2I_n^{\text{f.t.}} %= (-1)^{n-1}(1-2^{2-2n}) {2 \zeta_{2n-1}\over \pi^{2n-1}}
\,.
\end{align}
Here the octagon function $I_n^{\text{oct}}$ is evaluated for $\xi=y=0$. We will use this specific case throughout our analysis.~\footnote{ 
For arbitrary $\xi$ and $y$, the function $I_n^{\text{oct}}$ can be expressed in terms of ladder integrals \cite{Belitsky:2020qir}.}
The additional factor of $2$ in front of $I_n^{\text{f.t.}}$ in the last relation in \re{Is} is due to the double degeneracy of the zeros and poles of the functions  $\Phi_{\rm loc}(x)$ and $\Phi_{\rm oct}(x|0,0)$ defined in \re{cases}.

Finally, the last term in \re{F-gen} takes into account exponentially small, nonperturbative corrections to $\mathcal F(g)$.
As shown in  \cite{Basso:2009gh,Beccaria:2022ypy}, for the symbol function of a general form \re{chi-Phi},  the function $\Delta f(g)$ is given by a transseries that runs in powers of $e^{-8\pi g x_n}$  (with $n=1,2,\dots$) where $x_n$ are roots of the function \re{Phi}. In virtue of \re{xs} all roots of the symbols \re{chis} are proportional to the smallest root $x_1$. As a consequence, the function $\Delta f(g)$ takes simpler form 
\begin{align}\label{trans}
 \Delta f(g) = \sum_{n\ge 1} \left(g^{p} e^{-8 \pi g x_1} \right)^n \left[ A^{(n)}_{1}+\sum_{k=1}^{\infty}\frac{A^{(n)}_{k+1}}{2k(k+1)}g^{-k}\right],
\end{align}
where the expansion coefficients $A^{(n)}_{k+1}$ depend on $\chi(x)$ and $\ell$.
A nonnegative integer parameter $p$ is introduced in \re{trans} to account for the degeneracy of the roots of the functions  \re{cases}. 
In particular, $p=0$ for the symbols $\chi_{\text{W}}(x)$ and $\chi_{\text{f.t.}}(x)$.
For the symbols $\chi_{\rm loc}(x)$ and $\chi_{\rm oct}(x|0,0)$, the roots are double degenerate leading to $p=1$.

Applying the above relations, we can write the sum of the last two terms in \re{F-gen} as
\begin{align}\label{sum-f}
f(g) + \Delta f(g) {}& = \mathcal F_0(g) +\sum_{n\ge 1} \left(g^{p} e^{-8 \pi g x_1} \right)^n \mathcal F_n(g)\,,
\end{align}
where the coefficient functions are given by series in $1/g$
\begin{align}\label{goal}
\mathcal F_n(g)= A^{(n)}_{1}+\sum_{k=1}^{\infty}\frac{A^{(n)}_{k+1}}{2k(k+1)}g^{-k} \,.
\end{align}
The function $\mathcal F_0(g)$ takes the same form with $A^{(0)}_{1}=0$ and the expansion coefficients $A^{(0)}_{n}\equiv A_n$ given by \re{As}. The relation \re{sum-f} takes the expected form \re{ill}.~\footnote{For the sake of simplicity we did not include the first three terms in \re{F-gen} into the definition of the function $\mathcal F_0(g)$.}

Our goal in this paper is to develop a systematic method to compute the coefficient functions \re{goal} for $n=1,2,\dots$ and, then, investigate the resurgent properties of the strong coupling expansion \re{F-gen} and \re{sum-f}.

%A close examination reveals however that for all symbols in \re{chis} the expansion coefficients in \re{f} grow factorially $A_k \sim k!$ at large $k$.As a result, the perturbative expansion \re{f} suffers from Borel singularities. To make the strong expansion \re{F-gen} unambiguous,  it has to to supplemented with nonperturbative, exponentially small corrections. 
   
\subsection*{Truncated Bessel operator} 

The key to solving this problem lies in recognizing the deep connection between the interaction matrix \re{eq:K_nm} and
the truncated Bessel operator $\bm K_\chi$. 

This integral operator is defined as 
\begin{align}\label{Bes}
\bm K_\chi  \phi(x) = \int_{0}^{\infty}dy\, K(x,y)\chi\Big( {\sqrt{y}\over 2g}\Big)\phi(y) \,,
\end{align}
where $\phi(x)$ is a test function and $K(x,y)$ is given by an infinite sum of the normalized Bessel functions \re{psi} 
(hence the name of the operator)
\begin{align}\label{K(x,y)} \notag
  K(x,y) {}&=\sum_{n\ge 1} \psi_n(x)\psi_n(y) 
\\  {}& = {\sqrt x J_{\ell+1}(\sqrt x)J_{\ell}(\sqrt y) - \sqrt y J_{\ell+1}(\sqrt y)J_{\ell}(\sqrt x)\over 2(x-y)} 
 =\frac12 \int_0^1 dt \, t J_{\ell}(t\sqrt x) J_{\ell}(t\sqrt y) \,.
\end{align}
The relation on the second line follows from the properties of the Bessel function \cite{Tracy:1993xj}. 

As follows from \re{ortho}, the matrix elements of the integral operator \re{Bes} are identical to the entries of the semi-infinite matrix \re{eq:K_nm}  
\begin{align}
K_{nm}(g) = \langle \psi_n|\bm K_\chi|\psi_m \rangle = \int_{0}^{\infty}dx dy\, \psi_n(x) K(x,y)\chi\Big( {\sqrt{y}\over 2g}\Big)\psi_m(y)\,.
\end{align}
As a consequence, the determinant of the semi-infinite matrix in \re{F(g)} coincides with a Fredholm determinant of the operator \re{Bes} 
\begin{align}
\mathcal F(g) = \log \det(1- \bm K_\chi)\,.
\end{align}
This relation is powerful because it allows us to derive an exact relation for the function $\mathcal F(g)$ by exploiting the properties of the Bessel kernel \re{K(x,y)}.

As found in  \cite{Belitsky:2019fan,Belitsky:2020qrm,Belitsky:2020qir}, the function $\mathcal F(g)$ satisfies a differential equation    
\begin{align}\label{dF}
  \lr{g \partial_g }^2 \mathcal F(g) =4g^2\int_0^\infty dz\,z^2  q^2(z,g) \partial_z \chi(z)\,.
\end{align}
It involves an auxiliary function $q(z,g)$ which is defined as a matrix element of the resolvent of the truncated Bessel operator
\re{Bes}
\begin{align}\label{q}
q(z,g) = \langle x\vert\frac{1}{1-\bm{K}_{\chi}}\vert\phi_{0}\rangle\,,\qqqquad z={\sqrt x/(2g)}\,,
\end{align}
where $\phi_0(x) = J_\ell(\sqrt x)$ is a reference state. The $z-$variables were introduced in (\ref{q}) to simplify the analysis of (\ref{dF}) at strong coupling. 

Most importantly, we can use the properties of the Bessel kernel \re{Bes} to show that the function \re{q} satisfies a differential equation 
\begin{align}\label{eq:diffeq}
\Big((g\partial_{g})^{2} + 4g^{2}z^{2}-\ell^{2}+2g^2\partial^2_{g}\mathcal F(g)\Big) q(z,g)=0 \,.
\end{align}
It involves the second derivative of $\mathcal F(g)$ which, according to \re{dF}, depends on $q(z,g)$ in a nontrivial way. As a result, the
substitution of the solution of \re{eq:diffeq} into \re{dF} yields a complicated (integro-differential) equation for the function $\mathcal F(g)$. 

We would like to emphasize that the relations \re{dF} and \re{eq:diffeq} hold for any coupling constant $g$. Having computed the function $q(z,g)$, we can apply \re{dF} and determine $\mathcal F(g)$ up to a few integration constants. At weak coupling, these constants can be determined from \re{F-w}. At strong coupling, they follow from the asymptotic behaviour \re{F-gen}. 

To find the function $\mathcal F(g)$ at strong coupling we employ two different, complimentary techniques that we review below. 
 
\subsection*{Riemann-Hilbert problem}

To solve the system of coupled nonlinear equations \re{dF} and \re{eq:diffeq} at strong coupling, we need an appropriate ansatz for the function $q(z,g)$. As we show below, it can be derived by recasting the relation \re{q} into a Riemann-Hilbert problem for the function $q(z,g)$. 

We start with identifying analytical properties of the function $q(z,g)$. 
Using the completeness relations of the functions \re{psi} and  \re{ortho}, we can apply \re{q} to express $q(z,g)$ as a Neumann series of  
the Bessel functions
\begin{align}\label{q-N}
q(z,g) =\sum_{n\ge 1} c_n(g)\, \psi_n \lr{4g^2 z^2} \,,\qqquad c_n(g)=\bra{\psi_n }\frac{1}{1-\bm{K}_{\chi}}\vert\phi_{0}\rangle\,.
\end{align}
This series is assumed to be convergent for any $z$. According to their definition \re{psi}, the functions $\psi_n(4g^2z^2)$ are entire functions of $z$. This immediately implies that the function $q(z,g)$ is also an entire function of $z$ for any coupling constant $g$. 

Moreover, taking into account that $\psi_n \lr{4g^2 z^2}\sim J_{2n+\ell-1}(2gz)/z$, we can infer from \re{q-N} that $q(z,g)$ is a real-valued function of $z$ with a well-defined parity  
\begin{align}\label{q-parity}
q(z,g) = (-1)^\ell q(-z,g)\,,
\end{align}
where the nonnegative integer $\ell$ specifies the Bessel kernel \re{K(x,y)}. 

The function $q(z,g)$ satisfies an infinite set of integral equations  
\begin{align}\label{q-master}
\int_0^\infty dz\, (1-\chi (z))\, q(z,g) J_{2k+\ell-1}(2g z) =0\,,\qqqquad (k\ge 1)\,.
\end{align}
Indeed, changing the integration variable to $x=(2g z)^2$ and replacing $q(z,g)$ with its definition \re{q}, the integral in \re{q-master} can be expressed as
\begin{align}\label{zero}
\int_0^\infty dx\, \psi_k(x) \lr{1-\chi\Big( {\sqrt{x}\over 2g}\Big)} \langle x\vert\frac{1}{1-\bm{K}_{\chi}}\vert\phi_{0}\rangle
=\int_0^\infty dx\, \psi_k(x) \bra{x} \bm{K} \vert\phi_{0}\rangle = 0\,.
\end{align}
Here the integral operator $\bm{K}\equiv \bm{K}_{\chi=1}$ has the kernel \re{K(x,y)} and satisfies $\bm{K}\ket{\psi_k}=\ket{\psi_k}$ and $\bm{K}^2=\bm{K}$. The second relation in \re{zero} 
follows from the properties of the function $\phi_0(x)=J_\ell(\sqrt x)$.

The integral equations \re{q-master} have to be supplemented with the additional condition that their solution $q(z,g)$ has to be an entire function of $z$. The resulting Riemann-Hilbert problem can be solved using the technique developed in \cite{Beccaria:2022ypy}. We find that a general solution to \re{q-master} is given by
\begin{align}\label{q-a}
q(z,g) 
{}& ={1\over \sqrt{4\pi g b}} \left[e^{-{i\pi\over 2}\ell} (iz)^{-\beta-\frac12} {e^{2igz}\over \Phi(-z)} h_+(z,g) +e^{{i\pi\over 2}\ell}  (-iz)^{-\beta-\frac12} {e^{-2igz}\over\Phi(z)} h_-(z,g) \right].
\end{align}
It depends on the function $\Phi(x)$ and two parameters, $b$ and $\beta$, that enter the Wiener-Hopf decomposition of the symbol function \re{chi-Phi}.
The expression inside the brackets in \re{q-a} contains the sum of two oscillating terms which are accompanied by the coefficient functions $h_\pm(z,g)$. These functions are not independent of each other. Requiring \re{q-a} to satisfy the parity relation \re{q-parity}, leads to 
\begin{align}\label{h-par}
h_-(z,g) = h_+(-z,g)\,.
\end{align}
The solution to \re{q-master} is defined up to a multiplication of $q(z,g)$  by an arbitrary function of the coupling constant. The normalization factors in \re{q-a} were chosen for later convenience (see \re{f0-norm} below).

At weak coupling, the function $q(z,g)$ can be found from \re{q} by expanding the matrix element in powers of $\bm{K}_{\chi}$. Substitution of the resulting expression of $q(z,g)$ into \re{dF} yields the weak coupling expansion of $\mathcal F(g)$ that matches \re{F-weak} and \re{trB}.

At strong coupling, in a close analogy with \re{trans}, we can seek for the functions $h_\pm(z,g)$ in \re{q-a} in the form of a transseries 
\begin{align}\label{fpm-exp}
h_\pm(z,g)=h_\pm^{(0)}(z,g) + \sum_{n\ge 1} \lr{i e^{-8\pi g x_1}}^n h_\pm^{(n)}(z,g)\,,
\end{align}
where the coefficient functions $h_+^{(n)}(z,g)$ and $h_-^{(n)}(z,g) = h_+^{(n)}(-z,g)$ are given by series in $1/g$.
We emphasize that the relation \re{fpm-exp} holds for the symbol function of the form \re{chi-Phi} where the zeros $x_n$  are simple and proportional to $x_1$.~\footnote{For general values of the zeros $x_n$, the exponentially small corrections to \re{fpm-exp} run in powers of $e^{-8\pi gx_n}$. } For the symbol functions with degenerate zeros, like $\chi_{\text{loc}}(x)$ and $\chi_{\text{oct}}(x|0,0)$ in \re{chis}, we can slightly separate the zeros and take the limit of zero separation afterward.

Substituting the ansatz \re{q-a} and \re{fpm-exp} into the integral equation \re{q-master} and requiring $q(z,g)$ to be an entire function of $z$, we can derive  
nontrivial relations for the coefficient functions $h_\pm^{(n)}(z,g)$ (with $n=0,1,2,\dots$). The technique from \cite{Beccaria:2022ypy} allows for a straightforward computation of the first few terms of the expansion of these functions in $1/g$. However the calculation of higher-order terms becomes increasingly intricate.
 As we show below, the differential equation (\ref{eq:diffeq}) provides a powerful tool to tackle this problem. Namely, by requiring the function \re{q-a} to satisfy the differential equation \re{eq:diffeq} we will be able to determine the subleading coefficient functions $h_\pm^{(n)}(z,g)$ (with $n=1,2,\dots$) in terms of the leading function $h_\pm^{(0)}(z,g)$.

\subsection*{Method of differential equation}

To demonstrate a power of the differential equation \re{eq:diffeq}, we first apply it to reproduce the `perturbative' part of the function \re{F-gen} given by 
\re{f} and \re{As}. 

To this end, we
neglect exponentially small corrections to \re{fpm-exp} and substitute $h_\pm(z,g)=h_\pm^{(0)}(z,g)$ into \re{q-a}. 
This leads to
\begin{align}\label{q2}
q^2(z,g) = {z^{-2\beta-1}\over 2\pi g b\,\Phi(-z)\Phi(z)}  h_+^{(0)}(z,g) h_+^{(0)}(-z,g)    + O(e^{\pm 4 i g z})\,,
\end{align}
where the last term is a rapidly oscillating function of $z$. Plugging this relation into \re{dF} and recalling the definition of the symbol function 
\re{chi-Phi}, we arrive at
\begin{align}\label{ddF}
  \lr{g \partial_g }^2 \mathcal F(g) =-2g \int_0^\infty {dz\over\pi} \,z \partial_z \log(1-\chi(z))h_+^{(0)}(z,g) h_+^{(0)}(-z,g) +\dots\,,
\end{align}
Here dots denote the contribution of the last term in \re{q2}. We can show following \cite{Beccaria:2022ypy} that it is exponentially small at large $g$ and does not contribute to the perturbative part of $\mathcal F(g)$.

The relation \re{ddF} should be compared with the analogous relation
\begin{align}\label{ddF-exp}
  \lr{g \partial_g }^2 \mathcal F(g) =-g A_0+O(1/g) = -2g\int_{0}^{\infty}\frac{dz}{\pi} z\partial_{z}\log(1-\chi(z))+O(1/g)
\end{align}
that follows from \re{F-gen} and \re{A00}. Matching the leading $O(g)$ term in \re{ddF} and \re{ddF-exp} we find that
\begin{align}\label{f0-norm}
h_+^{(0)}(z,g)=1+O(1/g)
\end{align}
up to an overall sign. 

To find the subleading corrections to $h_+^{(0)}(z,g)$, we require that the function \re{q-a} has to satisfy the differential equation \re{eq:diffeq}. Replacing $\mathcal F(g)$ with its general expression \re{F-gen} and equating to zero the coefficients in front of powers of $1/g$ on the left-hand side of \re{eq:diffeq} we find  
\begin{align}\label{f0}
h_+^{(0)}(z,g) {}& = 1+\frac{i  (4 \ell _{\beta }^2-1 )}{16 z g}-\frac{(4 \ell _{\beta }^2-1)(4 \ell _{\beta }^2-9)+64 i z A_2}{512 z^2 g^2}
\\{}&   \notag
   -\frac{i \left((4 \ell _{\beta }^2-1)(4 \ell _{\beta }^2-9)(4 \ell _{\beta }^2-25)+192 i z A_2 (4 \ell _{\beta
   }^2-9)+2048 z^2 A_3\right)}{24576 z^3 g^3}+\dots,
\end{align}
where $\ell_\beta\equiv \ell+\beta$ and the coefficients $A_n$ are defined in \re{f}. Substituting \re{f0} into \re{ddF}, we obtain the expansion of
$\lr{g \partial_g }^2 \mathcal F(g)$ in powers of $1/g$. Its matching to \re{F-gen} leads to the system of equations for the coefficients $A_2,A_3,\dots$ whose solution is given by \re{As}.

It is straightforward to extend the above analysis to determine the nonperturbative, exponentially small correction to \re{fpm-exp}. To derive the leading coefficient function $h_+^{(1)}(z,g)$, we plug the ansatz \re{q-a} and \re{fpm-exp} into
the differential equation \re{eq:diffeq} and replace $\mathcal F(g)$ with its general expression \re{F-gen}, \re{f} and \re{trans}.
Equating to zero the coefficient of $e^{-8\pi g x_1}$ in \re{eq:diffeq} we get
\begin{align}\notag\label{f1}
{}& h_+^{(1)}(z,g) =-\frac{4 \pi x_1 A_1^{(1)}}{z+2 i \pi x_1}+\frac{\pi 
  x_1 (i (1-4 \ell ^2) A_1^{(1)}-4 z A_2^{(1)})}{4 z (z+2 i \pi 
  x_1) g}+{1\over 384 z^2 (z+2 i \pi x_1)^2 g^2}
\\{}& \notag 
  \times \Big[3 \pi x_1 \left(\left(4 \ell ^2-1\right) \left(z \left(4
   \ell ^2-25\right)+2 i \pi x_1 \left(4 \ell ^2-9\right)\right)+64 i z (z+2 i \pi
   x_1) A_2^{(0)}\right) A_1^{(1)}
 \\{}&  
   -8 z \left(3 \left(2 z^2+i \pi x_1 \left(4 \ell
   ^2-1\right) z+2 \pi ^2x_1^2 \left(1-4 \ell ^2\right)\right) A_2^{(1)}+16 \pi  z
  x_1 (z+2 i \pi x_1) A_{3}^{(0)}\right)\Big]+\dots
\end{align}
The coefficients of powers of $1/g$ are meromorphic functions of $z$ with poles located at $z=0$ and $z=-2\pi i x_1$. They depend on 
the `perturbative' coefficients $A_n^{(0)}\equiv A_n$ given by \re{As} and the coefficients $A_n^{(1)}$ defining the nonperturbative function \re{trans} for $p=0$,~\footnote{We assume for the moment that the zeros of the symbol function $1-\chi(x)$ are not degenerate. The case of degenerate roots is considered below.}    
\begin{align}\label{trans1}
 \Delta f(g) = \sum_{n\ge 1} \left(e^{-8 \pi g x_1} \right)^n \left[ A^{(n)}_{1}+\sum_{k=1}^{\infty}\frac{A^{(n)}_{k+1}}{2k(k+1)}g^{-k}\right].
\end{align}
%Repeating the calculation, 
In a similar manner, equating to zero the coefficient of $e^{-8\pi n g x_1}$ in \re{eq:diffeq}
we can derive the coefficient functions $h_+^{(n)}(z,g)$ for $n=2,3,\dots$ in terms of the $A^{(n)}-$coefficients. They are given by series in $1/g$ whose coefficients  have poles at $z=0$ and $z=-2\pi i m x_1$ (with $m=1,\dots,n$).  

As in the previous case, we can attempt to determine the coefficients $A_n^{(1)}$ by substituting the ansatz \re{q-a} into \re{dF} and matching the $O(e^{-8\pi gx_1})$ term on both sides of \re{dF}.
%\re{f1} into relations \re{q-a} and \re{fpm-exp}, and matching the $O(e^{-8\pi gx_1})$ term on both sides of \re{dF}. 
On the left-hand side of \re{dF}, this correction arises from the first term (for $n=1$) of the series \re{trans1}. However, on the right-hand side, it arises from two separate sources: from integrating rapidly oscillating terms of the perturbative function \re{q2} and from the $O(e^{-8\pi gx_1})$ term in $q^2(z,g)$, which itself originates from the leading nonperturbative correction \re{f1} to the function $q(z,g)$. This significantly increases the complexity of the calculation, especially at higher orders  in
$e^{-8\pi gx_1}$. 
   
We present below a more efficient method for determining the coefficients $A_n^{(1)}$ and, more generally, the nonperturbative function \re{trans}. This approach leverages the analytical properties of the function $q(z,g)$ defined in \re{q} and \re{q-a}. 

\subsection*{Quantization condition}

As explained above, the function $q(z,g)$ is an entire function of $z$ for any coupling $g$. 
We found that in the strong coupling limit, for $g \to \infty$ with $z$ fixed, it takes the form \re{q-a} with the coefficient functions given by \re{h-par}, \re{f0} and \re{f1}.  

Let us examine the analytical properties of \re{q-a}. Due to the presence of the functions $\Phi(\pm z)$ in the denominator, the two terms inside the brackets in \re{q-a} possess poles at $z = \pm 2\pi ix_n$ with $n = 1, 2, \dots$, (see \re{Phi}). The coefficient functions $h_\pm(z,g)$ introduce additional singularities in \re{q-a}.  As follows from \re{f0} and \re{f1}, these functions 
have poles at $z=0$ and $z= \pm 2\pi i n x_1$. The poles at the origin are artifacts of the large $g$ expansion. As was shown in \cite{Belitsky:2020qrm}, they can be regularized using an analytical regularization, for example. 

We are left over with the poles at $z = \pm 2\pi ix_n$ and $z = \pm 2\pi i n x_1$.
As was mentioned above, for all symbol functions defined in \re{chis} the set of roots $\{x_n\}_{n\ge 1}$ is always contained within the sequence $\{n x_1\}_{n\ge 1}$.~\footnote{It follows from \re{xs} that for the symbols
$\chi_{\text{f.t.}}(x)$ and $\chi_{\text{loc}}(x)$ the sequence $\{mx_1\}_{m\ge 1}$ includes $x_n=(2n-1)x_1$ for odd $m=2n-1$. Moreover, for the symbol $\chi_{\text{W}}(x)$ we have $x_1=1$ and $x_n=0$ for $n\ge 2$.} This implies that 
the poles at $z= \pm 2\pi i n x_1$ coincide with those of the functions $1/\Phi(\pm z)$ for $x_n=nx_1$ and are distinct otherwise. 

Thus, for general values of the $A-$coefficients in \re{f0} and \re{f1}, the function \re{q-a} exhibit poles at $z= \pm 2\pi i n x_1$ (for $n\ge 1$). However, since $q(z,g)$ is an entire function of $z$, the residues at these poles must be zero. This leads to the quantization condition for the  $A-$coefficients.
It can be expressed as
\begin{align}\label{qc}
\lim_{z\to -2\pi i n x_1} \lr{z+2\pi i nx_1} q(z,g) = 0 \,,\qqqquad (n\ge 1)\,.
\end{align}
On the shell of this relation,  the function $q(z,g)$ has vanishing residues at $z=2\pi i n x_1$ due to the parity property \re{q-parity}.

\subsection*{Summary of the method}

\begin{figure}
\begin{centering}
\includegraphics[width=14cm]{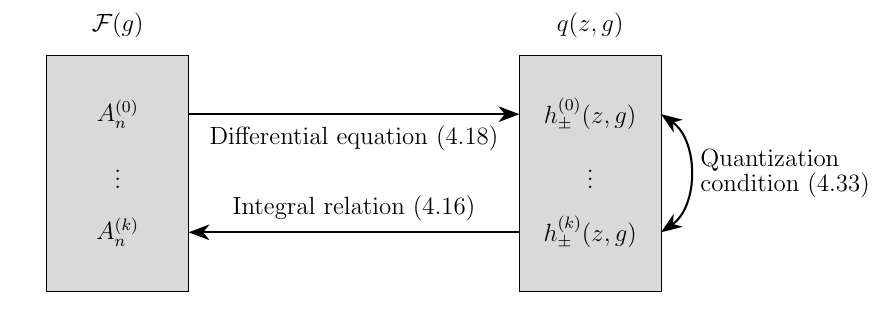}
\par\end{centering}
\caption{Schematic representation of the method for determining the strong coupling expansion \re{F-gen}. %of $\mathcal F(g)$.
 }
\label{method}
\end{figure}

The procedure of deriving the strong coupling expression of $\mathcal F(g)$ is outlined in Figure \ref{method}. 
We start by parameterizing $\mathcal F(g)$, as defined in \re{F-gen} and \re{sum-f}, using the coefficients $A^{(k)}_n$ introduced in \re{goal}. By substituting this ansatz into the differential equation \re{eq:diffeq} and replacing the function $q(z,g)$ with its general expression \re{q-a} and \re{fpm-exp}, we can express the coefficient functions  $h_{\pm}^{(k)}(z,g)$ in terms of the coefficients $A^{(k)}_n$. Imposing the quantization condition \re{qc} on the resulting expression for $q(z,g)$, allows us to express the coefficients $A^{(k)}_n$ for $k\ge 1$ in terms of the leading coefficients $A^{(0)}_n$. Finally, the latter coefficients can be determined using the integral relation \eqref{dF}, thereby yielding the strong coupling expansion \re{F-gen}. In the next section we elaborate on this procedure. 

\section{Nonperturbative corrections}\label{sect5}

We demonstrate in this section that the relation \re{qc} alone allows us to determine the nonperturbative function \re{trans1}.  

\subsection*{Leading order}

Let us start with solving \re{qc} for $n=1$. We choose $z=-2\pi i x_1+\varepsilon$ and examine the relation \re{q-a} for $\varepsilon\to 0$. 

It follows from \re{Phi} that the function $\Phi(z)$ vanishes in this limit,  $\Phi(z)= \varepsilon \, \Phi'(-2\pi i x_1)+O(\varepsilon^2)$, whereas $\Phi(-z)$ approaches a finite value $\Phi(2\pi i x_1)$. In a similar manner, one finds from \re{f1} that
$h_+(z,g)$ exhibits poles in $1/\varepsilon$ and $h_-(z,g)=h_+(-z,g)$ remains finite. 
Taking these relations into account, the equation \re{qc} can be expressed as
\begin{align}\label{l1}
\lim_{\varepsilon\to 0} \, \varepsilon \, h_+(-2\pi i x_1+\varepsilon,g) = 4\pi x_1 \Lambda_1 e^{-8\pi g x_1} h_+(2\pi i x_1,g) \,.
\end{align}
This relation implies that the function $h_+(-2\pi i x_1+\varepsilon,g)$ has a simple pole at $\varepsilon=0$.
% and fixes the residue at this pole. 
Here the notation was introduced for
\begin{align}\label{L1}
\Lambda_1=  {e^{i\pi \ell_\beta}\over 4\pi i x_1} {\Phi(2\pi i x_1)\over \Phi'(-2\pi i x_1)} \,,
\end{align}
where $\ell_\beta=\ell+\beta$ and $\Phi'(x)=\partial_x\Phi(x)$. 

Replacing the function $h_+(z,g)$ in \re{l1} with its transseries representation \re{fpm-exp} and comparing the coefficients of $e^{-8\pi g k x_1}$ on both sides of \re{l1},
we find a recurrence relation between the coefficient functions $h_+^{(k-1)}$ and $h_+^{(k)}$
\begin{align}\label{qc-lo}
\lim_{\varepsilon\to 0} \ \varepsilon \,  h_+^{(k)}(-2\pi i x_1+\varepsilon,g) = -4\pi i x_1 \Lambda_1 h_+^{(k-1)}(2\pi i x_1,g) \,,
\end{align}
where $k\ge 1$. 

For $k=1$ we can replace $h_+^{(0)}$ and $h_+^{(1)}$ in \re{qc-lo} by their expressions \re{f0} and \re{f1}, respectively. By comparing the coefficients of powers of $1/g$ on both sides of the resulting equation, we can then determine the $A^{(1)}-$coefficients in terms of the perturbative $A-$coefficients \re{As}
\begin{align}\notag\label{A1-coeffs}
{}& A_1^{(1)}= i \Lambda_1\,,
\\[1.2mm]\notag
{}& A_2^{(1)}= \frac{i \Lambda_1   (4 \ell _{\beta}^2-1)}{4 \pi  x_1}\,,
\\
{}& A_3^{(1)}= \frac{3 i \Lambda_1   ((4 \ell _{\beta}^2-1)(4 \ell _{\beta}^2-9)-64 \pi  A_2 x_1)}{128 \pi ^2 x_1^2}\,,\qquad \dots
\end{align}
The expression for $A_1^{(1)}$ agrees with the findings of \cite{Beccaria:2022ypy}. The relations \re{A1-coeffs} were obtained by solving \re{qc-lo} for $k=1$. 
We found that the equations \re{qc-lo} are automatically satisfied for $k\ge 2$ for the $A^{(1)}-$coefficients given by \re{A1-coeffs}.

We employ the relations \re{A1-coeffs} to compute the leading $O(e^{-8\pi g x_1})$ correction to the nonperturbative function \re{trans1}
\begin{align}
 \Delta f(g) =  e^{-8 \pi g x_1}  \left(A^{(1)}_{1}+\frac{A^{(1)}_2}{4 g}+\frac{A^{(1)}_3}{12 g^2}+\dots\right)+O(e^{-16\pi g x_1})\,.
\end{align}
Note that the coefficients \re{A1-coeffs} are proportional to $\Lambda_1$ and, furthermore, their dependence on parameters $\ell$ and $\beta$ enters solely through their sum $\ell_\beta=\ell+\beta$. The latter property also applies to the perturbative function \re{f} and \re{As}.

\subsection*{Beyond the leading order} 
 
Let us examine the quantization condition \re{qc} for $n\ge 2$. 
As before, we choose $z=-2\pi i n x_1+\varepsilon$ and replace $q(z,g)$ with its expression \re{q-a}.
Going to the limit $\varepsilon\to 0$ we get from \re{qc} 
\begin{align}\label{l2}
\lim_{\varepsilon\to 0} \varepsilon \, h_+(-2\pi i n x_1+\varepsilon,g) = 4\pi n x_1 \Lambda_n\, e^{-8\pi n g x_1} h_+(2\pi i n x_1,g) \,,
\end{align}
where the notation was introduced for
\begin{align}\label{La-n}
\Lambda_n = \lim_{\varepsilon\to 0}  {e^{i\pi \ell_\beta}\over 4\pi i n x_1} {\varepsilon\, \Phi(2\pi i n x_1)\over \Phi(-2\pi i n x_1+\varepsilon)} \,.
\end{align}
To evaluate the limit in this relation, we need to distinguish between two cases depending on whether
$\Phi(-2\pi i n x_1+\varepsilon)$ vanishes for $\varepsilon\to 0$ or not. This distinction is equivalent to whether 
$nx_1$  coincides with $x_n$:
\begin{itemize}
\item
For $nx_1=x_n$, or equivalently $ \Phi(-2\pi i n x_1)=0$ we get from \re{La-n}
\begin{align}\label{Ln}
\Lambda_n = {e^{i\pi \ell_\beta}\over 4\pi i n x_1} { \Phi(2\pi i n x_1)\over \Phi'(-2\pi i n x_1)} \,.
\end{align} 
\item
For $nx_1\neq x_n$, or equivalently $ \Phi(-2\pi i n x_1)\neq 0$, we have instead
\begin{align}\label{Ln-zero}
\Lambda_n =0 \,.
\end{align}
\end{itemize}
For $n=1$ the relation \re{Ln} coincides with \re{L1}.
 
By substituting \re{fpm-exp} into \re{l2}, we obtain the quantization condition
\begin{align}\label{qc-nlo}
\lim_{\varepsilon\to 0} \ \varepsilon \,  h_+^{(k)}(-2\pi i n x_1+\varepsilon,g) = 4\pi (-i)^{n} n x_1 \Lambda_n \, h_+^{(k-n)}(2\pi i n x_1,g) \,.
\end{align}
where $k\ge n$. This relation generalizes \re{qc-lo} to arbitrary $n\ge 1$. For $k=n$, the right-hand side of \re{qc-nlo} involves the perturbative function \re{f0}. 
The function $h_+^{(n)}(z)$ on the left-hand side of \re{qc-nlo}  can be computed using the differential equation \re{eq:diffeq} as explained in the previous section. In a close analogy with \re{f0} and \re{f1}, this function depends on the $A^{(m)}-$coefficients with $0\le m\le n$. 
Then, solving the equations \re{qc-nlo} for $k=n$ and $n=2,3,\dots$ we can determine the $A^{(n)}-$coefficients in terms of the perturbative coefficients \re{As}. 

In this way, we can compute the coefficients of $e^{-8\pi n g x_1}$ in the transseries \re{trans1} for any $n=2,3,\dots$.
For instance, at order $O(e^{-16\pi g x_1})$ the first few coefficients are
\begin{align}\notag\label{A2-coeffs}
A_1^{(2)} {}&= \frac{1}{2}(\Lambda _1^2+2 i \Lambda _2)\,,
 \\\notag
A_2^{(2)}{}&=\frac{(2 \Lambda _1^2+i \Lambda _2)  (4 \ell_\beta ^2-1)}{8 \pi  x_1}\,,
 \\ \notag
A_3^{(2)}{}&= \frac{3 \Lambda _1^2 ((4 \ell_\beta ^2-1)(4 \ell_\beta ^2-5)-32 \pi  A_2 x_1)}{64 \pi ^2 x_1^2} 
\\ {}& 
+ \frac{3 i \Lambda _2((4 \ell_\beta ^2-1)(4 \ell_\beta ^2-9)-128 \pi  A_2 x_1)}{512 \pi ^2 x_1^2} \,.
\end{align}
The coefficients at order  $O(e^{-24\pi g x_1})$ are
\begin{align}\notag\label{A3-coeffs}
A_1^{(3)}={}&\frac{1}{9} \left(-3 i \Lambda _1^3+8 \Lambda _2 \Lambda _1+9 i
   \Lambda _3\right)\,,
\\[2mm]\notag
A_2^{(3)}={}&\frac{\left(-3 i \Lambda _1^3+4 \Lambda _2 \Lambda _1+i
   \Lambda _3\right)  (4 \ell_\beta ^2-1)}{12 \pi  x_1}\,,
\\\notag
A_3^{(3)}={}& -\frac{3 i
   \Lambda _1^3 \left((4 \ell_\beta ^2-1)(12\ell_\beta ^2-11)-64 \pi  A_2 x_1\right)}{128 \pi ^2 x_1^2}
\\ \notag
 {}&+  \frac{ \Lambda _2
   \Lambda _1 (3(4 \ell_\beta ^2-1)(4 \ell_\beta ^2-5)-128 \pi  A_2 x_1)}{64 \pi ^2 x_1^2}
\\{}&
 +\frac{i \Lambda _3((4 \ell_\beta ^2-1)(4 \ell_\beta ^2-9)-192 \pi  A_2 x_1)}{384 \pi ^2 x_1^2}   \,.
\end{align}
These expressions depend on the perturbative coefficients \re{As} and the parameters $\Lambda_n$ defined in \re{Ln} and \re{Ln-zero}.
We verified that for the $A^{(n)}-$coefficients given by \re{A2-coeffs} and \re{A3-coeffs}, the relations \re{qc-nlo} are automatically satisfied for arbitrary $k>n$.

The above analysis readily generalizes to compute the $A^{(n)}-$coefficients for any $n$. 
To write down a general expression for $A_{k}^{(n)}$, it is convenient to assign weight $n$ to $\Lambda_n$. The product
$\Lambda_{n_1}\Lambda_{n_2}\dots$ has the weight $n_1+n_2+\dots$. As follows from \re{A2-coeffs} and \re{A3-coeffs}, the coefficients  $A_{k}^{(2)}$ and $A_{k}^{(3)}$ are given by the sum of terms, each with a homogenous weight $2$ and $3$, respectively. 
A general expression of $A_{k}^{(n)}$ is given by a multilinear combinations of the $\Lambda-$parameters defined in \re{La-n}  of homogenous weight $n$
\begin{align}\label{A-gen}
A_{k}^{(n)} = \sum_{m_1+2m_2+\dots+n m_n =n} (i\Lambda_{1})^{m_1}(i\Lambda_{2})^{m_2} \dots (i\Lambda_{n})^{m_n} \,P_{\bm{m}}(\ell_\beta^2)\,.
\end{align}
Here the sum runs over nonnegative integers $\bm{m}=\{m_1,\dots,m_n\}$ satisfying
  $0\le m_i \le n/i$. The expansion coefficients $P_{\bm{m}}(\ell_\beta^2)$ are polynomials of degree $(k-1)\ge 0$ in $\ell_\beta^2$. 
It is straightforward to compute them for any $n$ and $k$ by following the steps outlined above. 
 
By substituting the expressions derived above into relation \re{trans1}, we can calculate the nonperturbative function $\Delta f(g)$. Note that the coefficients \re{A-gen} are accompanied in \re{trans1} by the factor of $ e^{-8\pi g n x_1}=\prod_{p=1}^n e^{-8\pi g p m_p x_1}$. This effectively transforms the right-hand side of \re{trans1} into an expansion in powers of $(i \Lambda_p\, e^{-8\pi g p x_1})$.

In the next section, we present the explicit expressions for the nonperturbative function $\Delta f(g)$ evaluated for the symbol functions defined in \re{chis}.
 
\subsection*{Degenerate symbols}

In the above discussion we assumed that the symbol function $1-\chi(x)$ has simple zeros (see \re{chi-Phi}). This holds for $\chi_{\text{W}}(x)$ and  $\chi_{\text{f.t.}}(x)$ defined in \re{chis} but not for the functions $\chi_{\text{loc}}(x)$ and $\chi_{\text{oct}}(x|0,0)$. 
In the latter case, the zeros of $1-\chi(x)$ are double degenerate. When attempting to apply the above expressions for the $A^{(n)}-$coefficients we find that the corresponding $\Lambda-$parameters \re{L1} and \re{Ln} diverge because $\Phi'(-2\pi i n x_1)=0$. 

This highlights the distinction in the form of nonperturbative function for symbols with degenerate roots. For double degenerate roots, this function differs from \re{trans1} and looks as ~\cite{Beccaria:2022ypy}
\begin{align}\label{trans2}
 \Delta f(g) = \sum_{n\ge 1} \left(g e^{-8 \pi g x_1} \right)^n \left[ A^{(n)}_{1}+\sum_{k=1}^{\infty}\frac{A^{(n)}_{k+1}}{2k(k+1)}g^{-k}\right].
\end{align}
To determine the expansion coefficients, we can slightly split the degenerate roots apart, $x_n\to x_n^\pm=x_n \pm \delta$, perform the calculations following the steps described above and finally take the limit $\delta\to 0$.

Let us compute the leading $O(g e^{-8\pi gx_1})$ correction to \re{trans2}. To this end, we introduce the function
\begin{align}\label{Phi-delta}
 \Phi_\delta(x) = \lr{1-{ix\over 2\pi x_1^+}}\lr{1-{ix\over 2\pi x_1^-}}F(x)\,,
\end{align}
where $x_1^\pm=x_1\pm \delta$. For $\delta= 0$ it coincides with the function $\Phi_{\delta=0}(x)=\Phi(x)$ which appears in the Wiener-Hopf decomposition \re{chi-Phi} of the symbol function with a double degenerate root at $x=\pm 2\pi i x_1$.
For $\delta\neq 0$, the regularized function $ \Phi_\delta(x)$ has simple roots at $x=-2\pi i x_1^\pm$ and the previous analysis applies. However, an important difference is that we must now distinguish the nonperturbative corrections running  in powers of $e^{-8\pi g x_1^+}$ and $e^{-8\pi g x_1^-}$, % e.g.
\begin{align}\label{ans-delta}\notag
\Delta f(g) {}&= e^{-8 \pi g x_1^+} \left[ A^{(1),+}_{1}+\sum_{k=1}^{\infty}\frac{A^{(1),+}_{k+1}}{2k(k+1)}g^{-k}\right]
\\
{}& + e^{-8 \pi g x_1^-} \left[ A^{(1),-}_{1}+\sum_{k=1}^{\infty}\frac{A^{(1),-}_{k+1}}{2k(k+1)}g^{-k}\right]+\dots
\end{align}
where dots denote subleading exponentially small corrections. 

To determine the expansion coefficients in \re{ans-delta}, we apply the differential equation \re{eq:diffeq} and seek its solution in the form \re{q-a} where the function $h_+(z,g)$ is given by
\begin{align} 
h_+(z,g)=h_+^{(0)}(z,g) +  i e^{-8\pi g x_1^+} h_+^{(1),+}(z,g)+  i e^{-8\pi g x_1^-} h_+^{(1),-}(z,g)+\dots\,.
\end{align}
The perturbative function $h_+^{(0)}(z,g)$ is regular in the limit $\delta\to 0$. For $\delta=0$ it is given by \re{f0} with the coefficients $A_n$ defined in \re{As}. In a similar manner, the functions $h_+^{(1),\pm}(z,g)$ are given by \re{f1} in which $x_1$ and $A^{(n)}$ are replaced with $x_1^\pm=x_1\pm \delta$ and $A^{(n),\pm}$, respectively. 
Being an entire function, $q(z,g)$ has to satisfy \re{qc} with $x_1$ replaced by $x_1^\pm$
\begin{align}\label{qc1}
\lim_{z\to -2\pi i x_1^\pm} \lr{z+2\pi i x_1^\pm} q(z,g) = 0 \,.
\end{align}
As before, this relation can be used to determine the $A^{(1),\pm}-$coefficients. 

The resulting expressions for the $A^{(1),\pm}-$coefficients are similar to those in \re{A1-coeffs}
\begin{align}
A_1^{(1),\pm}= i \Lambda_1^\pm\,,
\qqqquad
A_2^{(1),\pm}= \frac{i \Lambda_1^\pm   (4 \ell _{\beta}^2-1)}{4 \pi  x_1^\pm }\,, \quad \dots
\end{align}
where $\Lambda_1^\pm$ is defined as
\begin{align}
\Lambda_1^\pm =  {e^{i\pi \ell_\beta}\over 4\pi i x_1^\pm} {\Phi_\delta(2\pi i x_1^\pm)\over \Phi_\delta'(-2\pi i x_1^\pm)} 
=\mp  {e^{i\pi \ell_\beta}\over \delta}  {x_1 F(2\pi ix_1)\over F(-2\pi ix_1)}+O(\delta^0)\,.
\end{align}
Here in the second relation we replaced $\Phi_\delta(x)$ with \re{Phi-delta}. Both $\Lambda_1^+$ and $\Lambda_1^-$
diverge for $\delta\to 0$ but, most importantly, their sum $\Lambda_1^++\Lambda_1^-$ remains finite in this limit.

Substituting the above relations into \re{ans-delta}, we find that both terms on the right-hand side of \re{ans-delta} diverge as $1/\delta$ but their sum approaches a finite value for $\delta\to 0$ 
\begin{align}\label{f-B}
\Delta f(g) = -2i x_1 {d\over dx_1}  \left[e^{-8\pi g x_1} \lr{B^{(1)}_{1}+\sum_{k=1}^{\infty}\frac{B^{(1)}_{k+1}}{2k(k+1)}g^{-k}}  \right]+ O(e^{-16\pi g x_1})\,.
\end{align}
The coefficient functions are given by 
\begin{align}\notag\label{B1-coeffs}
{}& B_1^{(1)}=  \widetilde\Lambda_1\,,
\\[1.2mm]\notag
{}& B_2^{(1)}= \frac{\widetilde\Lambda_1   (4 \ell _{\beta}^2-1)}{4 \pi  x_1}\,,
\\
{}& B_3^{(1)}= \frac{3 \widetilde\Lambda_1   ((4 \ell _{\beta}^2-1)(4 \ell _{\beta}^2-9)-64 \pi  A_2 x_1)}{128 \pi ^2 x_1^2}\,.
\end{align}
Here $A_2$ is the perturbative coefficient \re{As} and $\widetilde\Lambda_1$ is defined in terms of the function $\Phi(x)=\lim_{\delta\to 0}\Phi_\delta(x)$ and its second derivative %as
\begin{align}
\widetilde\Lambda_1=e^{i\pi \ell_\beta}{F(2\pi ix_1)\over F(-2\pi ix_1)}=  {e^{i\pi \ell_\beta}\, \Phi(2\pi ix_1)\over 2 (2\pi i x_1)^2\Phi''(-2\pi ix_1)} \,.
\end{align}
Note that the % coefficient functions in \re{f-B} depend on $x_1$.
%The 
relations \re{B1-coeffs} can be obtained from \re{A1-coeffs} by replacing $i\Lambda_1$ with $\widetilde\Lambda_1$.
 
The derivative on the right-hand side of \re{f-B} arises due to the degeneracy of the roots of the symbol function. It produces the additional factor of $g$ and enhances the leading nonperturbative correction in agreement with \re{trans2}. Matching \re{f-B} to \re{trans2} we can determine the coefficients of the strong coupling expansion of the nonperturbative function \re{trans2}
\begin{align}\notag\label{A-deg}
{}& A_1^{(1)} = 16 i\pi x_1 B_1^{(1)}\,,
\\[1.2mm]\notag
{}& A_2^{(1)} = 16 i\pi x_1 \lr{B_2^{(1)}-{1\over 2\pi} {d\over dx_1}B_1^{(1)}}\,,
\\
{}& A_{n\ge 3}^{(1)} = 16 i\pi x_1 \lr{B_n^{(1)}-{n \over 8\pi(n-2)} {d\over dx_1}B_{n-1}^{(1)}} \,.
\end{align}

The above analysis can be easily extended to determine the higher order coefficients in \re{trans2} for a generic symbol function \re{chi-Phi} with degenerate zeros. We do not give the general expressions for the $A^{(n)}-$coefficients to avoid clutter. Instead, we present below the explicit expressions of the non-perturbative functions in \re{trans2} for the specific cases of the symbols defined in \re{chis}.
   
\section{Applications}\label{sect6}  

In this section, we apply the technique described above to derive the strong coupling expansion of the function \re{F-gen} for the symbol functions \re{chis} that we encountered in supersymmetric Yang-Mills theories.

\subsection{Half-BPS Wilson loop}

The expectation value of the half-BPS Wilson loop in planar $\mathcal N=4$ SYM is given by \re{W-exp}.
At strong coupling, we apply the known properties of the Bessel function to find 
\begin{align}\notag\label{W-str}
\vev{W} =\sqrt{2\over\pi} (4\pi g)^{-3/2} e^{4\pi g} {}& \left[\left(1-\frac{3}{32 \pi  g}-\frac{15}{2048 \pi ^2 g^2}-\frac{105}{65536 \pi ^3
   g^3} +\dots\right)  \right.
   \\ {}& \left.
   -ie^{-8\pi g}\left(1+\frac{3}{32 \pi  g}-\frac{15}{2048 \pi ^2 g^2}+\frac{105}{65536 \pi ^3 g^3}+\dots\right) \right].
\end{align}
The `perturbative' series on the first line has factorially growing coefficients and suffers from Borel singularities. The non-perturbative term in the second line is necessary to ensure that the sum of the two terms is well defined.
 
As was mentioned above, the same result \re{W-exp} can be obtained by computing the determinant \re{F(g)} for $\ell=2$ and the symbol function $\chi_{\text{W}}(x)$ (see \re{chis}) 
%\begin{align}\label{W-F}
%\vev{W} = e^{\mathcal F_{\rm W}(g)} \,.
%\end{align}
Using \re{W-exp} and \re{W-str}, we can validate the strong coupling expansion \re{F-gen} and also understand the properties of $\mathcal F_{\rm W}(g)=\log \vev{W}$  for finite coupling.
Specifying the general expression \re{F-gen} for the symbol function $\chi_{\text{W}}(x)$, we get 
\begin{align} \label{F-W} 
{}& {\cal F}_{\text{W}}(g)=4\pi g-\frac{3}{2} \log (4\pi g)-\frac12 \log{\pi\over 2}+f_{\text{W}}(g)+\Delta f_{\text{W}}(g)\,,
\end{align}
where we replaced the leading coefficients $A_0$ and $A_1$ with their expressions \re{A0} and \re{A1} for $\ell=2$ and $\beta=-1$ (see \re{betas}). In addition, we used the result for the Widom-Dyson constant $B$ from \cite{Beccaria:2022ypy}.  
Applying \re{f}, \re{As} and \re{Is}, we find that the perturbative function \re{f} is given by  (for $\ell_\beta=\ell+\beta=1$)
\begin{align}\label{PT-W}
f_{\text{W}}(g)=-\frac{3}{32 \pi  g}-\frac{3}{256 \pi ^2 g^2}-\frac{21}{8192 \pi ^3 g^3}-\frac{27}{32768\pi ^4 g^4} +O(1/g^5)\,.
\end{align}
Neglecting the last term in \re{F-W} we verified that $e^{{\cal F}_{\text{W}}(g)}$ correctly reproduce the first line in \re{W-str}.

To calculate the nonperturbative function $\Delta f_{\text{W}}(g)$, we first identity the zeros of the function $1-\chi_{\text{W}}(x)=1+(2\pi)^2/x^2$. Obviously, it has a pair of simple roots located at $x=\pm 2\pi i x_1$ with $x_1=1$. The nonperturbative function $\Delta f_{\text{W}}(g)$ takes the form \re{trans1}. It involves the expansion coefficients defined in \re{A1-coeffs}, \re{A2-coeffs} and \re{A3-coeffs}. Replacing $\Phi(x)=1-ix/(2\pi)$ in \re{L1} and \re{Ln-zero}, we obtain the $\Lambda-$parameters as $\Lambda_1=-1$ and $\Lambda_n=0$ for $n\ge 2$.
The resulting expression of the nonperturbative function \re{trans1} is
\begin{align}\notag\label{nonPT-W}
\Delta f_{\text{W}}(g) {}&= e^{-8\pi g}\lr{ -i-\frac{3 i}{16 \pi  g}-\frac{9 i}{512 \pi ^2 g^2}-\frac{51 i}{8192 \pi ^3
   g^3}+\dots}
   \\ 
 {}& + e^{-16\pi g}\lr{  
   \frac{1}{2}+\frac{3}{16 \pi 
   g}+\frac{9}{256 \pi ^2 g^2}+\frac{39}{4096 \pi ^3
   g^3}+\dots}+ O(e^{-24\pi g})\,.
%\\
% {}&  +e^{-24\pi g}\lr{  
%   \frac{i}{3}+\frac{3 i}{16 \pi 
%   g}+\frac{27 i}{512 \pi ^2 g^2}+\frac{123 i}{8192 \pi ^3
%   g^3}+\dots } + O(e^{-32\pi g})\,.
\end{align}
As a consistency check, we verified that the substitution of \re{F-W}, \re{PT-W} and \re{nonPT-W} into $e^{{\cal F}_{\text{W}}(g)}$  yields the expected strong coupling expansion \re{W-str}.

The perturbative series in \re{PT-W} as well as the series accompanying powers of $e^{-8\pi g}$ in \re{nonPT-W} have factorially growing coefficients. 
The resurgence properties of these series follow from those of the Bessel function in \re{W-str}.
 
Unlike the strong coupling expansion \re{W-str}, which contains a single nonperturbative term proportional to $e^{-8\pi g}$, the nonperturbative function \re{nonPT-W} involves an arbitrary power of $e^{-8\pi g}$. The presence of the additional exponential terms in \re{nonPT-W} stems from working with the strong coupling expansion of
${\cal F}_{\text{W}}(g)=\log \vev{W}$ rather than $\vev{W}$. This suggests that, for a general symbol function, the strong coupling expansion of $e^{\mathcal F(g)}$ might be simpler than the expansion of $\mathcal F(g)$.

\subsection{Flux tube correlators}

As in the previous example, the symbol function $\chi_{\text {f.t.}}(x)$ defined in \re{chis} enables exact calculation of the corresponding function $\mathcal F_{\text{f.t.}}(g)$ for arbitrary coupling $g$. 
For $\ell=0$ it is given by~\cite{Beccaria:2022ypy}
\begin{align}\label{Fft-ex}
\mathcal F_{\text{f.t.}}(g) = \frac38 \log\cosh(2\pi g) -\frac18 \log{\sinh(2\pi g)\over 2\pi g}\,.
\end{align} 
Unlike the previous example, its weak coupling expansion has a finite radius of convergence $g_\star=\pm i/4$, or equivalently $\lambda_\star=-\pi^2$. It is  a solution to $\cosh(2\pi g_\star)=0$ closest to the origin.   

At strong coupling, we get from \re{Fft-ex}
\begin{align}\label{Fft-str}
\mathcal F_{\text{f.t.}}(g) = \frac{g\pi}2 + \frac18 \log\lr{g\pi\over 2} +\frac{1}{2}e^{-4\pi g}-\frac{1}{8}e^{-8\pi g}+\frac{1}{6}e^{-12\pi g}-\frac{1}{16}e^{-16\pi g}+\dots
\end{align}
This relation exhibits two intriguing properties: the perturbative series terminates at order $O(g^0)$, and the nonperturbative terms have trivial, $g-$independent coefficient functions.

These properties are readily apparent once we plug the symbol function $\chi_{\text {f.t.}}(x)$ into the general expression \re{F-gen}. Indeed, the perturbative coefficients in \re{As} are proportional to $4\ell_\beta^2-1$. Taking into account \re{betas} we find that  $\ell_\beta=\ell+\beta_{\text{f.t.}}=-1/2$ for $\ell=0$ and, therefore,   
the perturbative function \re{f} vanishes. We use \re{A0} and \re{A1} to verify that the first three terms on the right-hand side of \re{F-gen} match their counterparts in \re{Fft-str}.

The function $1-\chi_{\text {f.t.}}(x)=\coth(x/2)$ has simple zeros at $x=\pm 2\pi i x_n$ with $x_n=n-1/2$. The corresponding nonperturbative function $\Delta f_{\text{f.t.}}(g)$ is given by a general expression \re{trans1}. Examining the relations \re{A2-coeffs} and \re{A3-coeffs} we observe that for $\ell_\beta=-1/2$ all the expansion coefficients but 
$A_1^{(n)}$  vanish leading to
\begin{align}\label{ft-simp} 
 \Delta f_{\text{f.t.}}(g) = \sum_{n\ge 1}  e^{-4 \pi n g}  \, A^{(n)}_{1} \,.
\end{align}
To compute the coefficients in this relation, we first apply \re{Ln} and \re{Ln-zero} and replace the function $\Phi=\Phi_{\text{f.t.}}(x)$ with its expression \re{cases} to get 
\begin{align}
\Lambda_{2n-1} =-\frac{i}{2\pi}\left[{\Gamma (n-\frac{1}{2} )\over \Gamma(n)}\right]^2\,, \qqqquad  \Lambda_{2n}=0\,,
\end{align}
where $n\ge 1$.
Substituting these relations into \re{A1-coeffs},  \re{A2-coeffs} and  \re{A3-coeffs} we find that 
\begin{align}
A^{(n)}_{1}=\frac{1-3 (-1)^n}{8 n}\,.
\end{align}
We use this relation to verify that the function \re{ft-simp}  correctly reproduces the nonperturbative terms in \re{Fft-str}.
   
\subsection{Octagon}   

Unlike the previous examples, the function  $\mathcal F_{\text{oct}}(g)$ corresponding to the symbol function $\chi_{\rm oct}(x|y,\xi)$ defined in \re{chis} does not allow for a closed-form expression. 

At strong coupling, for $g\to\infty$ with the kinematical variables $y$ and $\xi$ held fixed, we can apply \re{F-gen} to derive the asymptotic expansion of $\mathcal F_{\text{oct}}(g)$. 
The perturbative part of the strong coupling expansion \re{F-gen} has been studied in details in \cite{Belitsky:2020qrm,Belitsky:2020qir,Beccaria:2022ypy} and we refer an interested reader to this paper for details. Here we shall concentrate on the nonperturbative function $\Delta f_{\rm oct}(g|y,\xi)$. Its properties depend on whether solutions to 
$1-\chi_{\rm oct}(x|y,\xi)=0$ are degenerate. Denoting solutions to this equation as $x=2\pi i x_n^\pm$ we find from \re{chis}
\begin{align}\label{x-xi}
x_n^\pm=n \sqrt{1\pm {i\xi\over \pi n} }\,.
\end{align}
For $\xi\neq 0$ the zeros $x_n^\pm$ are distinct and the nonperturbative function $\Delta f_{\rm oct}$ takes the form \re{ans-delta}.
For $\xi=0$ the zeros become double degenerate and the function  $\Delta f_{\rm oct}$ is given by \re{f-B}.

In what follows we calculate the nonperturbative function $\Delta f_{\rm oct}$ at the specific kinematic point $y=\xi=0$. This choice is motivated by two reasons:  
the nonperturbative corrections are enhanced by powers of $g$ due to the degeneracy of zeros \re{x-xi} and, in addition, the symbol functions $\chi_{\rm oct}(x|0,0)$ and $\chi_{\rm loc}(x)$ defining two different functions $\mathcal F_{\rm oct}(g)$ and $\mathcal F_{\rm loc}(g)$
are related to each other as (see \re{chi-rel})
\begin{align}\label{chi-chi}
1-\chi_{\rm oct}(x|0,0)={1\over 1-\chi_{\rm loc}(x)}\,. %=\tanh^2(x/2)\,.
\end{align}
We show below that this leads to an interesting relation between the functions $\mathcal F_{\rm oct}(g)$ and $\mathcal F_{\rm loc}(g)$. 

For $y=\xi=0$ the octagon function \re{F-gen} takes the form
\begin{align}\label{F-oct}
\mathcal F_{\rm oct}(g)=-\pi  g+\frac12 \log g +B_{\rm oct} + f_{\rm oct}(g) + \Delta f_{\rm oct}(g)\,,
\end{align}
where the first two terms on the right-hand side follow from \re{A0} and  \re{A1} for $\ell_{\rm oct}=0$ and $\beta_{\rm oct}=1$. The Tracy-Widom constant is given by \cite{Belitsky:2020qrm,Belitsky:2020qir}
\begin{align}\label{B-oct}
B_{\rm oct}= - 6\log{\sf A} +\frac12+{7\log 2\over 6} +\log\pi\,,
\end{align}
where ${\sf A}$ is the Glaisher's constant. 

The perturbative function in \re{F-oct} looks as
\begin{align}\label{f-oct}
f_{\rm oct}(g)={3\over 8} \log\lr{g'\over g}+\frac{15 \zeta (3)}{32 (4\pi g')^3}+\frac{945 \zeta
   (5)}{256 (4\pi g'')^5}-\frac{765 \zeta (3)^2}{64 (4\pi g')^6}  
   +\frac{637875 \zeta(7)}{8192 (4\pi g')^7}
   +O(1/g'{}^8)\,,
\end{align}
where $g'=g+(\log 2)/\pi$ and $\zeta(n)$ are Riemann zeta-values. It can be obtained from \re{f} upon replacing the expansion coefficients with their expressions \re{As} and \re{Is} for $\ell_\beta=1$. The primary motivation for changing the expansion parameter in \re{f-oct} to $g'$ is to eliminate the appearance of $\log 2$ at all orders.
%In AdS/CFT description, the The can be viewed as an additive renormalization of the  
 
The nonperturbative function in \re{F-oct} is given by a general expression \re{trans2}. Going through the calculation of the expansion coefficients \re{A-deg} for the octagon symbol function \re{cases}, we arrive at
\begin{align}\label{trans-oct}
\Delta f_{\rm oct}(g) = 16i(4\pi g') e^{-8\pi g'} f^{(1)}_{\rm oct}(g)
 + 128 (4\pi g')^2 e^{-16\pi g'} f^{(2)}_{\rm oct}(g)+O(e^{-24\pi g'})\,.
\end{align}
where we inserted the superscript in the coefficient functions to indicate the order in $e^{-8\pi g}$. In close analogy to \re{f-oct}, it is advantageous to use the shifted coupling constant $g'$ to avoid the appearance of $(\log 2)$ in the nonperturbative functions.
These functions are given by series in $1/g'$
\begin{align}\notag\label{f-np-oct}
f^{(1)}_{\rm oct}(g) {}& = 1-\frac{\frac{7}{4}}{ (4\pi g')}-\frac{\frac{63}{32}}{ (4\pi g')^2}+\frac{\frac{165}{128}}{(4\pi g')^3}-\frac{\frac{45 \zeta
   (3)}{16}+\frac{3405}{2048}}{(4\pi g')^4}+\dots
\\[2mm]
f^{(2)}_{\rm oct}(g) {}& =1-\frac{\frac{7}{2}-\frac{81 i}{4}}{(4\pi g')}-\frac{\frac{3}{4}+\frac{1431i}{32}}{(4\pi g')^2}+\frac{\frac{309}{32}-\frac{10287 i}{512}}{(4\pi g')^3}   
   +\frac{-\frac{45 \zeta
   (3)}{8}-\frac{537}{128}+\frac{26325
   i}{4096}}{(4\pi g')^4}
 +\dots
\end{align} 
Note that the expansion coefficients in $f^{(2)}_{\rm oct}$ have both real and imaginary parts. As we show below this property plays an important role in verifying the resurgence relations for the octagon function \re{F-oct}. 
 
\subsection{Localization}   

For the symbol function $\chi_{\rm loc}(x)$ in \re{chis}, the relation \re{F-gen} takes the form
\begin{align}\label{F-loc}
\mathcal F_{\rm loc}(g)=\pi  g-\frac32 \log g +B_{\rm loc} + f_{\rm loc}(g) + \Delta f_{\rm loc}(g)\,.
\end{align}
As before, the first two terms on the right-hand side can be obtained from \re{A0} and  \re{A1} for $\ell_{\rm loc}=2$ and $\beta_{\rm loc}=-1$. The Tracy-Widom constant is given by (see \cite{Beccaria:2022ypy})
\begin{align}
B_{\rm loc}=-6 \log {\sf A} +\frac{1}{2}-\frac{11 \log  2 }{6}\,.
\end{align}
It differs from the analogous expression \re{B-oct} for the octagon.

The perturbative function \re{f} is given by
\begin{align}\label{f-loc}
f_{\rm loc}(g)={3\over 8} \log\lr{g''\over g}-\frac{15 \zeta (3)}{32 (4\pi g'')^3}-\frac{945 \zeta
   (5)}{256 (4\pi g'')^5}-\frac{765 \zeta (3)^2}{64 (4\pi g'')^6}
   -\frac{637875 \zeta(7)}{8192 (4\pi g'')^7}
   +O(1/g''{}^8)\,,
\end{align} 
where the expansion coefficients are given by \re{As} and \re{Is} (evaluated at $\ell_\beta=1$). Similar to \re{f-oct}, all terms in \re{f-loc} proportional to $\log 2$ can absorbed into redefinition of the coupling constant $g''=g-(\log 2)/\pi$. Note that this coupling differs from $g'$ by the sign of the $(\log 2)/\pi$ term. The two coupling constants are related to each other as $g''\to -g'$ for $g\to -g$. 

Employing the techniques described in the previous section, we can calculate  the nonperturbative function $\Delta f_{\rm loc}(g)$. Applying the relations \re{trans2} and \re{A-deg} we get
\begin{align}\label{trans-loc}
\Delta f_{\rm loc}(g)=\frac{i}{32}(4\pi g'') e^{-4\pi g''} f^{(1)}_{\rm loc}(g)+{1\over 2048}(4\pi g'')^2 e^{-8\pi g''}  f^{(2)}_{\rm loc}(g)+O(e^{-12\pi g''})\,,
\end{align}   
where the coefficient functions are given by series in $1/g$. In a close analogy with \re{f-np-oct}, we expand them in powers of the shifted coupling constant $g''$
to get \footnote{Higher order terms of the transseries \re{f-np-oct} and \re{f-np-loc} can be found in an ancillary file associated with \cite{Bajnok:2024epf}.}
\begin{align}\notag\label{f-np-loc}
f^{(1)}_{\rm loc}(g){}&= 1+\frac{\frac{1}{2}+8 \log (2)}{(4\pi g'')}+\frac{12 \log
   (2)-\frac{15}{8}}{(4\pi g'')^2}+\frac{\frac{45}{16}-15 \log
   (2)}{(4\pi g'')^3}
\\\notag
{}&   
   +\frac{\frac{45 \zeta (3)}{8}-\frac{1005}{128}+\frac{75 \log
   (2)}{2}}{(4\pi g'')^4}+\frac{-\frac{585 \zeta (3)}{16}+45 \zeta (3) \log
   (2)+\frac{7875}{256}-\frac{2205 \log
   (2)}{16}}{(4\pi g'')^5}+\dots 
\\[2mm]\notag
f^{(2)}_{\rm loc}(g){}&=1+\frac{1+16 \log (2)}{(4\pi g'')}+\frac{-3+64 \log ^2(2)+32 \log (2)}{(4\pi g'')^2}
\\ 
{}& 
+\frac{\frac{21}{4}+192 \log ^2(2)-48 \log (2)}{(4\pi g'')^3}+\frac{\frac{45 \zeta (3)}{4}-\frac{105}{8}-96 \log
   ^2(2)+60 \log (2)}{(4\pi g'')^4} + \dots
\end{align}
Unlike \re{f-np-oct}, the expansion coefficients in both series are real and explicitly depend on $\log 2$.

As previously mentioned, the strong coupling expansion of $e^{\mathcal F_{\text{loc}}(g)}$ is expected to be simpler than that of $\mathcal F_{\text{loc}}(g)$. Indeed, we use \re{F-loc} and \re{trans-loc} to find that the nonperturbative correction to $e^{\mathcal F_{\text{loc}}(g)}$ is given by
\begin{align}\notag
{}&e^{\Delta f_{\rm loc}(g)}=1+\frac{i}{32}(4\pi g'') e^{-4\pi g''} f^{(1)}_{\rm loc}(g)
\\
{}& \qquad +{1\over 4096}e^{-8\pi g''} \left[1+{3\over (4\pi g'')}-{\frac{15}{2}\over (4\pi g'')^2}+\frac{{135\over 4}}{(4\pi g'')^3}+\frac{\frac{45 \zeta
   (3)}{4}-\frac{1575}{8}}{(4\pi g'')^4}+\dots\right]+O(e^{-12\pi g''})\,.
\end{align}
We observe that, in distinction to \re{trans-loc}, the coefficient function of the 
subleading $O(e^{-8\pi g''})$ correction scales as $O((4\pi g'')^0)$ and does not contain $\log 2$.

Comparing the perturbative functions \re{f-oct} and \re{f-loc}, we observe that their expansion coefficients coincide up to a sign. Moreover, expanding the functions 
$f_{\rm loc}(g)$ and $f_{\rm oct}(-g)$ as power series in $1/g$, one can check that they coincide order by order leading to
\begin{align}\label{f=f}
f_{\rm loc}(g)=f_{\rm oct}(-g)\,.
\end{align}
This relation is rather formal because the functions on both
sides  suffer from Borel singularities. It should be
understood as an equality between the expansion coefficients in the two series. Interestingly, the leading $O(g)$ terms in \re{F-oct} and \re{F-loc} exhibit the same relationship under the transformation $g\to -g$. However, this relation does not extend to the $O(\log g)$ terms and to the $B-$constants. 

As explained in \cite{Beccaria:2022ypy}, the relation \re{f=f} follows from the analogous relation between the symbol functions \re{chi-chi}. It arises because the perturbative function defined in \re{f} and \re{As} is formally invariant under the transformation $g\to -g$ and $I_n\to -I_n$. As follows from \re{prof}, the latter transformation corresponds to $\log(1-\chi(x))\to -\log(1-\chi(x))$. It is equivalent to exchanging the zeros, $x_n$,  and poles, $y_n$, of the symbol function \re{chi-Phi}.

A natural question arises: can the relation \re{f=f} be extended to the nonperturbative functions \re{trans-oct} and \re{trans-loc}?
Naive substitution $g \to -g$ in the transseries \re{trans-oct} and \re{trans-loc} yields a nonsensical result because it transforms exponentially small terms into exponentially large ones. As we will show in the next section, the underlying reason is that the strong coupling expansion of the functions $\mathcal F_{\text{oct}}(g)$ and $\mathcal F_{\text{loc}}(g)$ is affected by the Stokes phenomenon. In other words, the asymptotic expansions \re{trans-oct} and \re{trans-loc} are only valid for large positive values of the coupling constant $g$. They undergo a significant transformation after the substitution $g\to -g$ when $g$ becomes large negative.

We argue below that the two different transseries defining the strong coupling expansion of the functions $\mathcal F_{\text{oct}}(g)$ and $\mathcal F_{\text{loc}}(g)$ can be viewed as the asymptotic expansion of the \textit{same} function $\mathcal F_{\text{uni}}(g) $ for large negative and positive values of $g$, respectively. This function is defined piecewise as 
\begin{align}\label{F}
\mathcal F_{\text{uni}}(g) = \left\{\begin{array}{ll} 
f_{\rm loc}(g)+\Delta f_{\rm loc}(g)\,, & \qquad g>0 
\\[4mm]  f_{\rm oct}(-g)+\Delta f_{\rm oct}(-g)\,, & \qquad g<0 \end{array}\right.  
\end{align}
It has a Stokes line at ${\rm Re} \, g=0$, where its asymptotic expansion dramatically changes form. 
    
\section{Resurgence relations}\label{sect7}   

In this section we study the properties of the transseries that appear in the strong coupling expansion of the function \re{F-gen}. 

Among the four functions analyzed in the previous section, only $\mathcal F_{\text{f.t.}}(g)$ enjoys a remarkably simple strong coupling expansion (\ref{Fft-str}) due to its closed-form representation (\ref{Fft-ex}).  
In particular, its perturbative expansion does not contain $O(1/g)$ terms and the nonperturbative terms have trivial coefficient functions. These properties do not hold for the remaining functions $\mathcal F_{\text{W}}(g)$, $\mathcal F_{\text{oct}}(g)$ and $\mathcal F_{\text{loc}}(g)$. 
 
The strong coupling expansion of the function $\mathcal F_{\text{W}}(g)$ shares properties with the expansion of the Bessel function
(see \re{W-exp} and \re{W-str}). In particular, the perturbative and nonperturbative functions, defined in \re{PT-W} and \re{nonPT-W}, respectively, involve asymptotic series in $1/g$ with coefficients that grow factorially. While both functions suffer from Borel singularities,  ambiguities are cancelled out in their sum $f_{\text{W}}(g)+\Delta f_{\text{W}}(g)$. 
We show below that, despite the absence of a closed-form representation in terms of known special functions, the same property holds true for the functions $\mathcal F_{\text{oct}}(g)$ and $\mathcal F_{\text{loc}}(g)$ given by \re{F-oct} and \re{F-loc}, respectively.
  
\subsection*{Borel transform} 
  
The perturbative functions \re{f-oct} and \re{f-loc}  are defined as series in $1/g$, e.g.
\begin{align}\label{f-ser} 
f_{\text{oct}}(g) = \sum_{n\ge 0} {\alpha_n\over (4\pi g)^n} \,,
\end{align}
where the expansion coefficients $\alpha_n$ can be read from \re{f-oct}. In virtue of \re{f=f}, the expansion coefficients
of $f_{\text{loc}}(g)$ are given by $(-1)^n \alpha_n$. Due to the factorial growth of the expansion coefficients at large orders, $\alpha_n\sim n!$, these asymptotic series can be regarded as merely formal series. In order to give meaning to them, we use Borel summation.

We thus define the Borel transform
\begin{align}\label{B-def} 
\mathcal B_{\text{oct}}(s) = \sum_{n\ge 0}\alpha_n {s^n\over\Gamma(n+1)}\,.
\end{align}
The function $\mathcal B_{\text{loc}}(s)$ is given in the same way with $\alpha_n$ replaced by $(-1)^n \alpha_n$.
As a consequence, it is related to $\mathcal B_{\text{oct}}(s)$ as
\begin{align}\label{B-rel}
\mathcal B(s)\equiv \mathcal B_{\text{loc}}(s) = \mathcal B_{\text{oct}}(-s) \,.
\end{align}
The factorial growth of the expansion coefficients $\alpha_n$ (see \re{alpha} below) are transformed to constant asymptotics for the coefficients of the Borel transform $\mathcal B(s)$, implying a finite radius of convergence of the series \re{B-def}. 
The Borel transform develops singularities at the boundary of the convergence region. As we show below, they are located on the real axis for both positive and negative values of $s$.

The function $f_{\text{oct}}(g)$ is given by the inverse of the Borel transform which involves the integration of the function 
$\mathcal B(s)$ for real positive $s$. To avoid the singularities of this function, it is costumary to define the lateral Borel resummations $S^+$ (or $S^-$) by slightly shifting the integration contour above (or below) the real axis, respectively,
\begin{align}\notag\label{f-Borel}
{}& S^{\pm}[f_{\text{oct}}](g) = 4\pi g\int_0^{\infty e^{\pm i\epsilon}}ds\, e^{-4\pi g s} \,\mathcal B(s)\,,
\\
{}& S^{\pm }[f_{\text{loc}}](g) = 4\pi g\int_0^{\infty e^{\pm i\epsilon}}ds\, e^{-4\pi g s} \,\mathcal B(-s)\,.
\end{align}
The singularities of ${\mathcal B}(s)$ at positive and negative $s$ impact $S^{\pm}[f_{\text{oct}}](g)$ and $S^{\pm }[f_{\text{loc}}](g)$, respectively.  

Henceforth, unless otherwise indicated, a formal power series $f(g)$ is understood as a laterally Borel resummed 
series $S^+[f](g)$. For convenience, we will write $f(g)$ instead of $S^+[f](g)$. This implies to the perturbative series \re{f-ser} 
as well as to the series $f_{\text{oct}}^{(n)}$ and $f_{\text{loc}}^{(n)}$ that enter the nonperturbative functions \re{trans-oct} and \re{trans-loc}.

Following \re{F} we can interpret the relations \re{f-Borel} as defining two separate branches of the perturbative contribution, $f(g)$, to the function $\mathcal F_{\text{uni}}(g)=f(g)+\Delta f(g)$. For arbitrary coupling $g$, its Borel transformation is given by
\begin{align}\label{f-int-B}
f(g) = \int_0^{\infty e^{i\epsilon}}ds\, e^{-s} \,\mathcal B\Big({s\over 4\pi g}\Big)\,.
\end{align}
For positive and negative values of $g$, it coincides with $f_{\text{oct}}(g)$ and $f_{\text{loc}}(g)$, respectively.  

The Borel singularities of the function \re{F-gen} were studied in \cite{Beccaria:2022ypy}. 
It was shown there that for a general symbol function $\chi(x)$ of the form \re{chi-Phi}, the Borel transform $\mathcal B(s)$ has singularities on the real axis located at 
$s=2x_n$ and $s=-2y_n$ where positive $x_n$ and $y_n$ parameterize simple zeros and poles of the symbol function \re{chi-Phi}. 
The freedom in choosing the integration contour in \re{f-Borel} reveals an intrinsic ambiguity in defining the perturbative function \re{f-int-B}. 

This ambiguity is proportional to the discontinuity of the Borel transform $\mathcal B(s/(4\pi g))$ across the positive semi-axis $s>0$. For positive and negative coupling $g$, it comes from the cuts $\mathcal B(s/(4\pi g))$  that start at $s=8\pi g x_n$ and $s=-8\pi g y_n$, respectively. At strong coupling, the leading contribution comes from the cuts at $s=8\pi g x_1$ and $s=-8\pi g y_1$ which are closest to the origin. It scales as $O(e^{-8\pi g x_1})$ for $g>0$ and $O(e^{8\pi g y_1})$ for $g<0$. 
The nonperturbative function $\Delta f(g)$ should have the same behaviour in order for the sum of the two functions $f(g)+\Delta f(g)$ to be well defined. Note that $f(g)$ and $\Delta f(g)$ depend separately on the lateral resummation we choose, which is $S^+$ by default. If we were using the $S^-$ resummation instead, we should use $S^-[f+(\Delta f)^*](g)=S^+[f+\Delta f](g)$. 

We recall that the zeros and poles of the symbol functions $\chi_{\rm oct}(g)$ and $\chi_{\rm loc}(g)$ are double degenerate. As a consequence, the corresponding
nonperturbative functions \re{trans-oct} and \re{trans-loc} involve terms of the form  $(g e^{-8\pi g})^n$ and $(g e^{-4\pi g})^n$, respectively, for arbitrary $n\ge 1$. These terms should match the contribution to \re{f-int-B} from the singularities of the Borel transform. This leads to the following general expression for the function 
$\mathcal B(s)$ 
\begin{align}\notag\label{B-lo}
\mathcal B(s) {}&= {a_0\over 1+s} + {a_1\over (2+s)^2 } +{a_2\over (3+s)^3} + \dots
\\
{}& +{b_0\over 2-s} + {b_1\over (4-s)^2 } +{b_2\over (6-s)^3} + \dots\,,
\end{align}
The first and second lines are given by the sum over poles of degree $n=1,2,\dots$ located at $s=-n$ and $s=2n$, respectively. The dots on both lines of \re{B-lo} denote poles for higher $n$ as well as terms containing subleading (logarithmic) singularities. These terms are specified below in \re{eq:B0s}.
Substituting \re{B-lo} into \re{f-int-B} we find that, in virtue of \re{B-rel}, the poles of \re{B-lo} on the first and second line contribute, respectively, to the functions $f_{\rm loc}(g)$ and $f_{\rm oct}(g)$.

\subsection*{Median resummation}

Let us recall  the relations between three seemingly unrelated objects \cite{Dorigoni:2014hea,Aniceto:2018uik}:
\begin{itemize}
\item The large-order behavior of the perturbative coefficients \re{B-def}
\item The analytical properties of the Borel transform \re{B-lo}
\item The well-definedness of the transseries \re{sum-f}
\end{itemize}
\noindent
For convenience, we concentrate on the two observables defined in \re{F}. 

The singularities of the Borel transform \re{B-lo} are in one-to-one correspondence with large order behaviour of the perturbative coefficients in \re{B-def}. The expansion coefficients $\alpha_n$ grow factorially at large $n$
%and their behaviour at large $n$ can be parameterize as
\begin{align}\notag\label{alpha}
\alpha_n= (-1)^n\sum_{k\ge 0} \bigg[ a_k^{(1)} \Gamma(n+1-k)+a_k^{(2)} {\Gamma(n+2-k)\over 2^{n+2-k}}+a_k^{(3)} {\Gamma(n+3-k)\over 3^{n+3-k}}+ \dots\bigg]
\\
+\sum_{k\ge 0} \bigg[ b_k^{(1)} {\Gamma(n+1-k)\over 2^{n+1-k}}+b_k^{(2)} {\Gamma(n+2-k)\over 4^{n+2-k}}+b_k^{(3)} {\Gamma(n+3-k)\over 6^{n+3-k}}+ \dots\bigg]\,.
\end{align}
where $a_k^{(p)}$ and $b_k^{(p)}$ (with $p\ge 1$) are real-valued coefficients independent of $n$.
Substituting \re{alpha} into \re{B-def}, we find that the terms proportional to $a_k^{(p)}$ and $b_k^{(p)}$ produce singularities of the Borel transform $\mathcal B(s)$ at $s=-p$ and $s=2p$, respectively.  

The nature
of these singularities depends on $k$. The terms in \re{alpha} with $k=0$ give rise to \re{B-lo} (with $a_0=a_0^{(1)}$ and $b_0=b_0^{(1)}$) and produce the poles of $\mathcal B(s)$.  The remaining terms in \re{alpha} with $k\ge 1$ contribute to the subleading (logarithmic) singularities of the Borel transform  
\begin{align}\label{eq:B0s}\notag
\mathcal B(s) {}&= \frac{b_{0}^{(1)}}{2-s}-\log(2-s) \sum_{k\ge 0} b_{k+1}^{(1)}\frac{(s-2)^{k}}{k!} 
\\
 {}& +\frac{b_{0}^{(2)}}{(4-s)^{2}}+\frac{b_{1}^{(2)}}{4-s}-\log(4-s) \sum_{k\ge 0}b_{k+2}^{(2)}\frac{(s-4)^{k}}{k!} +\dots \,,
\end{align}
where dots denote analogous terms containing singularities at negative $s=-1,-2,\dots $ and positive $s=6,8,\dots$.

When performing the lateral Borel resummation $S^+$ in \re{f-int-B}, the integration contour goes slightly above the poles and  the logarithmic cuts. This ensures a finite and well-defined result but leads to an imaginary part of $S^+[f](g)$. If we chose to integrate below the cut, we would get the complex conjugate result $S^-[f](g)$. The difference between the two expressions is related to the discontinuity of ${\mathcal B}(s)$ along the real axis 
\begin{align}\notag\label{d1}
S^+[f](g)-S^-[f](g) &=4\pi g\int_0^\infty ds\, e^{-4\pi g s} \left[ \mathcal B(s+i0)-  \mathcal B(s-i0)\right]
\\
{}&=2i\pi  (4\pi g\, e^{-8\pi g}) \left[ b_0^{(1)} +{b_1^{(1)}\over 4\pi g} + {b_2^{(1)} \over (4\pi g)^2 } +\dots  \right] +O(e^{-16\pi g})\,.
\end{align}
Since $S^+[f](g)$ and $S^-[f](g)$ are complex conjugates, this relation implies that the imaginary part of $S^+[f](g)$ is equal to the discontinuity \re{d1} divided by two.

In a similar manner, the sign alternating terms on the first line of \re{alpha} generate singularities of $\mathcal B(s)$ at negative $s$ and produce the discontinuity in the Borel resummations of the function $f(-g)$ for $g>0$
\begin{align}\label{d2} 
S^+[f](-g)-S^-[f](-g) =2i\pi   (4\pi g\, e^{-4\pi g}) \left[ a_0^{(1)} +{a_1^{(1)}\over 4\pi g} + {a_2^{(1)} \over (4\pi g)^2 } +\dots  \right]+O(e^{-8\pi g})\,.
\end{align}
We observe that the transseries on the right-hand side of \re{d1} and \re{d2} have the same form as those in nonperturbative functions \re{trans-oct} and \re{trans-loc}, respectively. This is to be expected, given that $\mathcal F_{\text{uni}}(g)=f(g)+\Delta f(g)$ should be a well-defined function of the coupling and, therefore, half of the discontinuities  \re{d1} and \re{d2} should cancel out in the sum of two functions.~\footnote{Recall that the functions $f(g)$ and $\Delta f(g)$ are defined using the latteral Borel summation $S^+$.}

The coefficient functions in \re{d1} and \re{d2} develop Borel singularities due to a factorial growth of the expansion coefficients. For instance, for the leading term in \re{d1}  and \re{d2}  we have at large $k$~\footnote{Here a superscript in $\tilde a_m^{(n)}$ and $\tilde b_m^{(n)}$ 
refers to the singularity of the Borel transform at $s=2n$.}
\begin{align}\label{c-c}\notag
a_{k+1}^{(1)}=\sum_{m\geq0}\left[\tilde a_{m}^{(2)}\,{\Gamma(k+2-m)} +\tilde a_{m}^{(3)}\frac{\Gamma(k+3-m)}{2^{k+3-m}}+\dots\right],
\\
b_{k+1}^{(1)}=\sum_{m\geq0}\left[\tilde b_{m}^{(2)}\,\frac{\Gamma(k+2-m)}{2^{k+2-m}}+\tilde b_{m}^{(3)}\frac{\Gamma(k+3-m)}{4^{k+3-m}}+\dots\right],
\end{align}
where $\tilde a_m^{(n)}$ and $\tilde b_m^{(n)}$ are real coefficients independent of $k$. Plugging the last expression into \re{eq:B0s}, we find that the series 
involving the coefficients $b_{k}^{(1)}$ produces the additional singularities of $\mathcal B(s)$  
\begin{equation}\label{add-sing}
\sum_{k\ge 0}b_{k+1}^{(1)}\frac{(s-2)^{k}}{k!}=\frac{\tilde b_{0}^{(2)}}{(4-s)^2}+\frac{\tilde b_{1}^{(2)}}{4-s}-\log(4-s)
\sum_{m\ge 0} \tilde b_{m+2}^{(2)}\frac{(s-4)^{m}}{m!} +\dots\,,
\end{equation}
where dots denote singularities at $s=6,8,\dots$. 

Thus, the singularities of the Borel transform \re{eq:B0s} at $s=4$ come from two different sources: from the subleading  perturbative coefficients $b_{k}^{(2)}$ in \re{alpha} at fixed $k$ and from large$-k$ asymptotic behaviour of the leading perturbative coefficients $b_{k}^{(1)}$. Their contribution to $\mathcal B(s)$ is different in nature. As follows from \re{eq:B0s} and \re{add-sing}, the contribution proportional to $b_{k}^{(2)}$ is real, whereas the contribution proportional to $\tilde b_k^{(2)}$ is purely  imaginary due to the factor of $\log(2-s)$ in the second term in \re{eq:B0s} evaluated slightly above the cut at $s=4$. 
 
Having determined the Borel transfrom $\mathcal B(s)$, we can apply \re{f-int-B} to compute the perturbative function $f(g)$. 
This function alone cannot describe the observable $\mathcal F_{\text{uni}}(g)$ because, as explained above, it has a discontinuity along the real axis and takes complex values for real $g$. To avoid these issues, the perturbative function $f(g)$ should be supplemented with a nonperturbative correction $\Delta f(g)$, which should be such that the sum $\mathcal F_{\text{uni}}(g)=f(g)+\Delta f(g)$ is a real well-defined function of the coupling $g$. 
For $g>0$ it looks as %a transseries %$\Delta f(g)$ 
\begin{align}\label{D-f}\notag
{}& \Delta f(g)=(4\pi ge^{-8\pi g})\sum_{k=0}{f_{k,+}^{(1)}\over (4\pi g)^k}+(4\pi ge^{-8\pi g})^{2}\sum_{k=0}{f_{k,+}^{(2)}\over(4\pi g)^k} +O(e^{-24\pi g})\,,
\\
{}& \Delta f(-g)=(4\pi ge^{-4\pi g})\sum_{k=0}{f_{k,-}^{(1)}\over (4\pi g)^k}+(4\pi ge^{-4\pi g})^{2}\sum_{k=0}{f_{k,-}^{(2)}\over(4\pi g)^k} +O(e^{-12\pi g})\,.
\end{align}
The minimal ansatz for these functions verifying the conditions mentioned above can be worked out using  the \emph{median resummation} \cite{Dorigoni:2014hea,Aniceto:2018bis}. 

It yields a prediction for the nonperturbative functions \re{D-f} in terms of the discontinuity of the perturbative functions \re{d1} and \re{d2}. In particular, it allows us to express 
the expansion coefficients in \re{D-f} it terms of various coefficients defined in \re{alpha} and \re{c-c} 
\begin{align}\label{f-r}\notag
{}& f_{k,+}^{(1)}=-i\pi b_{k}^{(1)}\,,\qqqquad f_{k,+}^{(2)}=-i\pi b_k^{(2)}-{\pi^2\over 2}\tilde b_{k}^{(2)}\,,
\\[2mm]
{}& f_{k,-}^{(1)}=-i\pi a_{k}^{(1)}\,,\qqqquad f_{k,-}^{(2)}=-i\pi a_k^{(2)}-{\pi^2\over 2}\tilde a_{k}^{(2)}\,.
\end{align}The leading non-perturbative terms in \re{D-f} involving $f_{k,\pm}^{(1)}$  merely
kill the imaginary contribution to $f(g)$ and $f(-g)$ coming from  logarithmic cuts $\mathcal B(s) \sim \log(2-s)$ and $\mathcal B(s) \sim \log(1+s)$, respectively. The subleading
non-perturbative terms $f_{k,\pm }^{(2)}$ not only takes care of
the cut-contribution  $\mathcal B(s)\sim \log(4-s)$ and $\mathcal B(s)\sim \log(2+s)$, but  also
cancel the imaginary part of the Borel transform of the first series in \re{D-f},
which itself has a logarithmic cut with coefficients $\tilde b_{k}^{(2)}$ and $\tilde a_{k}^{(2)}$.

\subsection*{Numerical checks}

The median resummation yields a definite prediction \re{f-r} for the nonperturbative coefficient functions entering the transseries \re{D-f}. 
In this subsection, we verify the relations \re{f-r} by computing numerically linear combinations of the perturbative coefficients on the right-hand side of \re{f-r} and
comparing them with the analytical results for the analogous nonperturbative coefficients derived in the previous section (see \re{trans-oct} and \re{trans-loc}).

In the previous section, we used the method of differential equations to compute the first few perturbative terms  \re{f-oct} and \re{f-loc} of the strong coupling expansion of $f_{\rm oct}(g)$ and $f_{\rm loc}(g)$, respectively. %We demonstrated that to the two functions are related to each other as \re{f=f}. 
Their analytical expressions are lengthy at high orders in $1/g$, and thus are not particularly useful for analyzing the large-order behavior of the expansion coefficients. It is advantageous to perform high-precision numerical calculations from the beginning. Another simplification is to switch to the coupling $g'=g-I_{1}/2$ and $g''=g+I_{1}/2$ for $f_{\rm oct}(g)$ and $f_{\rm loc}(g)$, respectively. This eliminates all terms in both functions proportional to $\log(2)$, thereby simplifying the expressions.

\begin{figure}
\begin{centering}
\includegraphics[width=14cm]{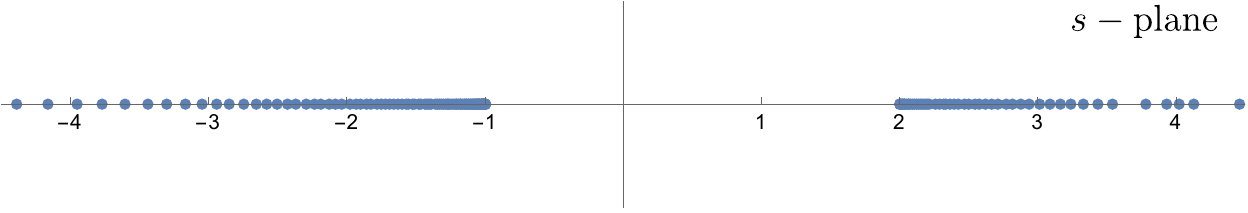}
\par\end{centering}
\caption{Poles of the diagonal Pade approximant $\mathcal B_{\text{app}}(s)$
in the case of $f_{\text{oct}}(g)$.}

\label{Pade}
\end{figure}

In the following, with a slight abuse of notation \re{f-ser}, we denote by $\alpha_n$ the expansion coefficients of $f_{\rm oct}(g)$ in $1/(4\pi g')$.  The expansion coefficients of $f_{\rm loc}(g)$ in $1/(4\pi g'')$ are given by $(-1)^n \alpha_n$. At large $n$, the coefficients $\alpha_n$ have the asymptotic behaviour \re{alpha} and are expected to satisfy the resurgence relations \re{f-r}. The nonperturbative coefficients $f_{k,+}^{(m)}$ and $f_{k,-}^{(m)}$ in these relations can be 
read as coefficients of $e^{-8\pi g' m}/{(4\pi g')}^{k-m}$ and $e^{-8\pi g'' m}/{(4\pi g'')}^{k-m}$ in \re{trans-oct} and \re{trans-loc}, respectively.

We applied the method of differential equation to compute the first $400$ coefficients $\alpha_n$ with $3000$ digits precision. We then considered a partial sum $\sum_{n=0}^{400}\alpha_{n} {s^{n}}/{\Gamma(n+1)}$ and constructed the diagonal $[200/200]$ Pad\'e approximant of the Borel transform \re{B-def} which we denote as $\mathcal B_{\text{app}}(s)$. The poles of the resulting expression of $\mathcal B_{\text{app}}(s)$ are shown in Figure \ref{Pade}.
It clearly indicates that the nearest singularities to the origin are at  $s=-1$ and $s=2$. The condensation of the poles on the real axis is in agreement with the expected asymptotical behavior \re{eq:B0s}.

\paragraph{Cut at $s=-1$:}
According to \re{alpha}, the leading asymptotic behaviour of the expansion coefficients is
\begin{align}
\alpha_n={\Gamma(n+1)\over (4\pi)^n}  (-1)^n\lr{a_0^{(1)}+{a_1^{(1)}\over n} +{a_2^{(1)}\over n(n-1)}+\dots} +
%2^{-n-1}\lr{b^{(1)}_0+{2b^{(1)}_1\over n} +{4b^{(1)}_2\over n(n-1)}+\dots}
\dots\,.
\end{align}
To extract the expansion coefficients in this relation, we applied high order Richardson extrapolation \cite{bender1999advanced} to the sequence of numerical values of $ (-1)^n \alpha_n(4\pi)^n/\Gamma(n+1) $ for $0\le n\le 400$ and obtained
\begin{align}\notag\label{a-coefs}
{}& a_{0}^{(1)} =-\frac{1}{32\pi}\,, && \frac{a_{1}^{(1)}}{a_{0}^{(1)}}=\frac{1}{2}+8\log(2)\,, 
\\\notag
{}& 
\frac{a_{2}^{(1)}}{a_{0}^{(1)}}=-\frac{15}{8}+12\log(2)\,,&& \frac{a_{3}^{(1)}}{a_{0}^{(1)}}=\frac{45}{16}-15\log(2)\,,
\\
{}& \frac{a_{4}^{(1)}}{a_{0}^{(1)}} =\frac{45\zeta(3)}{8}-\frac{1005}{128}+\frac{75\log(2)}{2}\,,&& \frac{a_{5}^{(1)}}{a_{0}^{(1)}}=-\frac{585\zeta(3)}{16}+45\zeta(3)\log(2)+\frac{7875}{256}-\frac{2205\log(2)}{16}  ,
\end{align}
with $56,54,\dots,44$ digit precision. 

\paragraph{Cut at $s=2$:}
To determine the $b^{(1)}-$coefficients parameterizing the cut of the Borel transform at $s=2$ we reexpanded the Pade approximant $\mathcal B_{\text{app}}(s)$ around $s=1/2$. The function $\mathcal B_{\text{app}}(s+1/2)$ has 
the cuts  at $(-3/2)$ and $3/2$. Repeating the asymptotic analysis, we obtained the $b_{k}^{(1)}$ coefficients as
\begin{equation}\label{b-coefs}
b_{0}^{(1)}=-\frac{16}{\pi}\,, \qqqquad\frac{b_{1}^{(1)}}{b_{0}^{(1)}}=-\frac{7}{4}\,, \qqqquad\frac{b_{2}^{(1)}}{b_{0}^{(1)}}=-\frac{63}{32}\,, \qqqquad\frac{b_{3}^{(1)}}{b_{0}^{(1)}}=\frac{165}{128}
\end{equation}
with $32,\dots,24$ digits precision. This precision was not enough to identify the coefficient of $\zeta(3)$ in the expression for the coefficient ${b_{4}^{(1)}}/{b_{0}^{(1)}} $ at the next order (see \re{b4} below).

A more efficient method for computing the coefficients in \re{alpha} is to exploit the generalized Borel transform
\begin{equation}\label{eq:genBoreltr}
\mathcal B_{p}(s)=\sum_{n=0}^{\infty}\frac{\alpha_{n}}{\Gamma(n+p+1)}s^{n+p}\,,
\end{equation}
where $p\ge 0$ is arbitrary parameter. For $p=0$ it coincides with \re{B-def}. The function $\mathcal B_{p}(s)$ has singularities at the same values of $s$ as $\mathcal B(s)$ but their nature depends on $p$. 
For instance, for $p=1/2$ we substitute \re{alpha} into \re{eq:genBoreltr} to find that $\mathcal B_{1\over 2}(s)$ has a square root cut at $s=2$
\begin{align}\label{B-cut}
\mathcal B_{\frac{1}{2}}(s) &=\frac{\sqrt{\pi}}{\sqrt{2-s}}\sum_{k\ge 0} \frac{b_{k}^{(1)}\sqrt{\pi}}{\Gamma\lr{k+\frac{1}{2}}}(s-2)^{k}+\dots\,,
\end{align}
where dots denote terms analytical at $s=2$. The simplest way to extract the coefficients $b_{k}^{(1)}$ is to open up the square root cut by introducing the new variable $s=2-z^{2}$ \cite{Costin:2020pcj}
\begin{align}\label{B-z2}
 {\mathcal B}_{\frac{1}{2}}(2-z^{2})=\sqrt\pi\sum_{k\ge 0} \frac{b_{k}^{(1)}\sqrt{\pi}}{\Gamma\lr{k+\frac{1}{2}}}(-1)^{k}z^{2k-1}+\dots\,,
\end{align}
where dots involve even powers of $z$. Expanding ${\mathcal B}_{\frac{1}{2}}(2-z^{2})$ around the origin, 
we can extract the $b_{k}^{(1)}$ coefficients from the terms with odd powers of $z$. 

As in the previous case, we use the numerical values of the first $400$ coefficients $\alpha_n$ to define a partial sum $s^{1/2}\sum_{n=0}^{400}\alpha_{n} {s^{n}}/{\Gamma(n+{3/2})}$ and, then, prepare its diagonal $[200/200]$ Pad\'e approximant which we denote as ${\mathcal B}_{\text{app},\frac{1}{2}}(s)$. According to \re{B-cut}, it has a square root cut at $s=2$.
Following the method explained above, we open up this cut by introducing $s(w)=2-(1-w)^{2}$. The cut at $s=2$ corresponds to $w=1$. Nevertheless, it is not possible to expand ${\mathcal B}_{\text{app},\frac{1}{2}}(s(w))$ around $w=1$ because
the Pad\'e approximant is only reliable away from its singularities. To circumvent  this issue, we expanded ${\mathcal B}_{\text{app},\frac{1}{2}}(s(w))$
around $w=0$, prepared its diagonal Pad\'e approximant and, then, expanded it in powers of $z=1-w$. 
By applying \re{B-z2}, we can be read off the coefficients 
$b_{n}^{(1)}$ from the terms involving odd powers of $z$. In this way, we reproduced the relations \re{b-coefs} and 
also computed two more coefficients
\begin{equation}\label{b4}
\frac{b_{4}^{(1)}}{b_{0}^{(1)}}=-\frac{45}{16}\zeta (3)-\frac{3405}{2048}\,,\qqqquad\frac{b_{5}^{(1)}}{b_{0}^{(1)}}=\frac{945}{64}\zeta (3)+\frac{25515}{8192}\,,
\end{equation}
with $33$ and $30$ digits precision, respectively. 

Repeating the same analysis for $s(w)=-1+(1-w)^2/2$, we reproduced the value of the coefficients \re{a-coefs} that paramerize the behaviour of the Borel transform in the vicinity of the cut at $s=-1$.

 We verify that, in agreement with the resurgence relations \re{f-r}, the ratios ${a_{n}^{(1)}}/{a_{0}^{(1)}}$ and ${b_{n}^{(1)}}/{b_{0}^{(1)}}$ coincide with the expansion coefficients of the nonperturbative function $f^{(1)}_{\rm loc}(g)$ and $f^{(1)}_{\rm oct}(g)$ defined in \re{f-np-loc} and \re{f-np-oct}, respectively. 

\paragraph{Cut at $s=4$:}
Let us examine the cut of the generalized Borel transform $\mathcal B_{\frac{1}{2}}(s)$ at $s=4$. Substituting \re{alpha} into \re{eq:genBoreltr} and taking into account \re{c-c}, we get for $s\to 4$
\begin{align}\label{B-cut4}
\mathcal B_{\frac{1}{2}}(s) {}& = \frac{\sqrt{\pi}(b_0^{(2)}+i\pi \tilde b_0^{(2)})}{2(4-s)^{3/2}}+\frac{\sqrt{\pi}}{\sqrt{4-s}} \sum_{k\ge 0}\frac{(b_{k+1}^{(2)}+i\pi \tilde b^{(2)}_{k+1}) \sqrt{\pi}}{\Gamma(k+\frac{1}{2})}(s-4 )^{k} +\dots \,,
\end{align}
where dots denote terms analytical at $s=4$. Here the terms proportional to $\tilde b^{(2)}_{k}$ result from \re{B-cut} after replacing the coefficients $b_k^{(1)}$ with their behavior \re{c-c} at large $k$. Note that the sum in \re{B-cut4} involves a linear combination of the same coefficients as in \re{f-r}, but their relative factor is different. 
 
 We can try to repeat the previous analyses by opening the cut at $s=4$ in \re{B-cut4} with an appropriate change of variables. Unfortunately, the cut at $s=2$ prevents us from doing so. In order to disentangle the two cuts, we use a conformal mapping
\begin{align}\label{conf}
s(z)=\frac{8z}{3(1+z^{2})-2z}\,.
\end{align}
It maps the two cuts at  $s=2$ and $s=4$  to $z=1$ and $z=z_4= (2+i\sqrt{5})/{3}$, respectively. The advantage of the conformal mapping \re{conf} is that it
opens up the square root cut  at $s=2$ and maps the cut at $s=4$ to the square root cut at $z=z_4$,
\begin{align}
s(z)=2-\frac32(z-1)^2 + O((z-1)^3)\,,\qqqquad s(z)=4+O(z-z_4)\,.
\end{align}
We can resolve the cut at $z=z_4$ by changing the variable in \re{conf} as $z(w)=z_4(1-w^2)$. Expanding the resulting expression for the Borel transform ${\mathcal B}_{\frac{1}{2}}(s(z(w)))$ around $w=0$, we can extract linear combination of the coefficients
$b_{k}^{(2)}+i\pi \tilde b^{(2)}_{k}$ by retaining the terms with odd powers of $w$. Taking its real and imaginary parts we can obtain the coefficients $b_{k}^{(2)}$ and $\tilde b^{(2)}_{k}$, respectively.

We used the previously defined Pad\'e approximant  ${\mathcal B}_{\text{app},\frac{1}{2}}(s)$ and proceeded as follows. We first expanded ${\mathcal B}_{\text{app},\frac{1}{2}}(s(z(w)))$  around $w=1/\sqrt 2$, prepared its diagonal Pad\'e approximant and, then, expanded it again around $w=0$. Comparing the terms with odd powers of $w$ with the analogous terms arising from the small $w$ expansion of the function ${\mathcal B}_{\frac{1}{2}}(s(z(w)))$ given by \re{B-cut4}, we 
can determine the numerical values of the $b_{k}^{(2)}$ and $\tilde b_{k}^{(2)}$ coefficients.  In this way, we obtained 
\begin{align}
 b_0^{(2)}=0\,, \qqqquad  \tilde b_0^{(2)}=-{256\over\pi^2}\,,
\end{align}
together with 
\begin{align}\notag\label{numbers}
{}&   {2 b_1^{(2)}\over \pi \tilde b_0^{(2)}}=\frac{81}{4}\,, &&  {2b_2^{(2)}\over \pi \tilde b_0^{(2)}}=-\frac{1431}{32}
,&& {2 b_3^{(2)}\over \pi \tilde b_0^{(2)}}=-\frac{10287}{512}\,,&& {2 b_4^{(2)}\over \pi \tilde b_0^{(2)}}=\frac{26325}{4096}\,,
\\[2mm]
{}&   {\tilde b_1^{(2)}\over \tilde b_0^{(2)}}=-\frac{7}{2}\,, && {\tilde b_2^{(2)}\over \tilde b_0^{(2)}}=-\frac{3}{4}, && \frac{\tilde b_{3}^{(2)}}{\tilde b_{0}^{(2)}}=\frac{309}{32}\,,&& \frac{\tilde b_{4}^{(2)}}{\tilde b_{0}^{(2)}}=-\frac{537}{128}-\frac{45\zeta(3)}{8}\,,
\end{align}
for $8,6,5,4$ digits precision.~\footnote{This precision was insufficient for identifying the numbers \re{numbers} unambiguously, but their
comparison to the exact values of the coefficients revealed satisfactory agreement.}
 
 \paragraph{Cut at $s=-2$:}
The above analysis can be repeated to describe the cut of the generalized Borel transform $\mathcal B_{\frac{1}{2}}(s)$ in the vicinity of the cut at $s=-2$. To disentangle the cuts at $s=-1$ and $s=-2$, we apply the conformal mapping $s(z)=- {4z}/\lr{3(1+z^{2})-2z}$ and go through the same steps as before. An important difference as compared with the previous case is that the expansion coefficients $a_k^{(2)}$ in \re{alpha} vanish  for $k\ge 0$.   
As a consequence, the behaviour of $\mathcal B_{\frac{1}{2}}(s)$ around $s=-2$ only depends on the coefficients $\tilde a_k^{(2)}$ defined in \re{c-c} . Going through the calculation we obtained
\begin{align}
{}& \tilde a_0^{(2)}=-{1\over 1024\pi^2}\,, && \quad {\tilde a_1^{(2)}\over \tilde a_0^{(2)}}=1+16 \log (2)\,, && \quad {\tilde a_2^{(2)}\over \tilde a_0^{(2)}}=64 \log ^2(2)+32 \log (2)-3\,.
\end{align}
 
 As in the previous case, we verify that, in agreement with the resurgence relations \re{f-r},  the ratios ${\tilde a_n^{(2)}/\tilde a_0^{(2)}}$ and ${\tilde b_n^{(2)}/\tilde b_0^{(2)}}+2ib_n^{(2)}/(\pi \tilde b_0^{(2)})$ coincide with the expansion coefficients of the nonperturbative functions $f^{(2)}_{\rm loc}(g)$ and $f^{(2)}_{\rm oct}(g)$ defined in \re{f-np-loc} and \re{f-np-oct}, respectively. 
    
 The results presented in this section provide a strong support of our findings. They also demonstrate that the median resummation correctly reproduces the nonperturbative functions \re{trans-oct} and \re{trans-loc} via the resurgence relations \re{f-r}. These relations establish a correspondence between nonperturbative, exponentially suppressed corrections and the asymptotic behavior of the perturbative series.

\section{Conclusions} 

In this paper we studied a special class of observables in $\mathcal N=2$ and $\mathcal N=4$ superconformal Yang-Mills theories which for an arbitrary 't Hooft coupling constant $\lambda$  admit representation as determinants of certain semi-infinite matrices. It is remarkable that similar determinants have previously appeared in the study of level-spacing distributions in matrix models and they are generalizations of the celebrated Tracy-Widom distribution. 

We exploited this relation to develop an efficient method for computing the observables in superconformal Yang-Mills theories at both weak and strong coupling. 
The weak coupling expansion is given by a series in 't Hooft coupling $\lambda$ which has a finite radius of convergence. At the same time, the strong coupling expansion is given by a series in $1/\sqrt\lambda$ with factorially growing coefficients. It suffers from Borel singularities 
necessitating the introduction of exponentially small corrections to render the strong coupling expansion well-defined and unambiguous. Being combined together, the `perturbative' corrections in $1/\sqrt\lambda$ and the `nonperturbative', exponentially suppressed corrections form a transseries. It involves an infinite sum of exponentially small terms, each accompanied by a series in $1/\sqrt\lambda$.  We computed explicitly the
expansion coefficients of these series and demonstrated that they are uniquely determined by large order behaviour of the expansion coefficients of the perturbative part of the transseries via the resurgence relations.
 
\begin{figure}
\begin{centering}
\includegraphics[width=12cm]{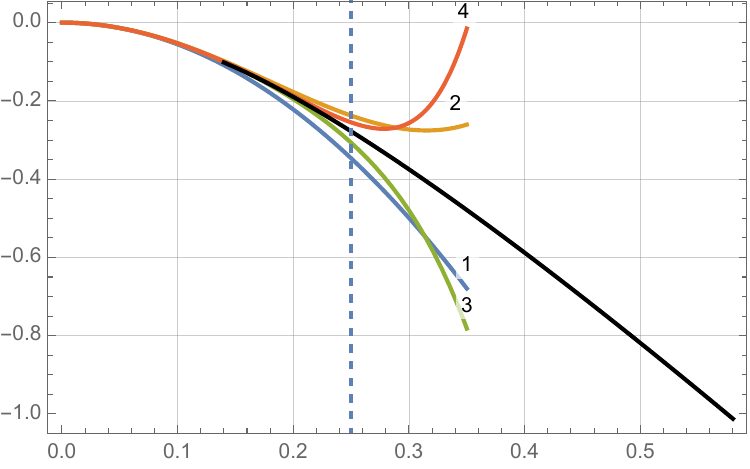}
\par\end{centering}
\caption{Dependence of the octagon function $\mathcal F_{\text{oct}}(g)$ on the coupling constant $g=\sqrt\lambda/(4\pi)$.  Colored curves starting at the origin represent the first few terms in the weak coupling expansion, integers indicate the number of terms retained. The black curve describes the strong coupling expansion defined by the latteral Borel summation. The vertical dashed line indicates the convergence radius of the weak coupling expansion.}
\label{plots}
\end{figure}

One may wonder what sense it makes to study non-perturbative corrections to observables, knowing in advance that they are exponentially small in the strong coupling regime. In addition to being needed to compensate Borel ambiguities in the strong coupling expansion, they play a very important role in the transition region between the strong and weak coupling regimes.  
Indeed, the non-perturbative corrections are of the same order of magnitude as the perturbative corrections for $\sqrt\lambda=O(1)$ and must therefore be taken into account.
By computing a sufficiently large number of perturbative and nonperturbative terms, we can use Borel summation to compute the observable for sufficiently small values of the coupling constant, within the convergence radius of the weak coupling expansion. This allows for the computation of observables for any value of the coupling constant. As an example, we show in Figure~\ref{plots} the dependence of the octagon function $\mathcal F_{\text{oct}}(g)$ on the coupling constant.

Another interesting result of our analysis is that non-perturbative corrections to two different observables (the four-point correlation  function of infinitely-heavy operators in planar $\mathcal N=4$ SYM at the special kinematical point and the free energy in $\mathcal N=2$ SYM defined on a unit sphere) are described by the same universal function defined for positive and negative values of the coupling constant $g=\sqrt\lambda/(4\pi)$.
This relation is surprising given that the two observables have different physical meaning and were obtained using different techniques -- integrability in planar $\mathcal N=4$ and localization in $\mathcal N=2$ SYM theories. It would be interesting to elucidate the underlying reason behind this phenomenon. 

 It is even more surprising that all observables described in the Introduction admit a representation as Fredholm determinants of the same truncated Bessel operator. The choice of the observable only determines the form of the cut-off (symbol) function. The Bessel operator is a representative of a large class of integrable Fredholm operators~\cite{Tracy:1993ah,Harnad:1997hf}. One can speculate that these operators emerge due to the integrability of the underlying gauge theory.  However, the precise reason for the appearance of the Bessel operator in four-dimensional gauge theories remains unclear and warrants further investigation.

Finding non-perturbative corrections at strong coupling in supersymmetric Yang-Mills theories and revealing their resurgence structure are notoriously difficult problems. So far, there are very few examples of such investigations in $\mathcal N = 4$ SYM \cite{Basso:2009gh,Aniceto:2015rua,Dorigoni:2015dha,Arutyunov:2016etw, Dorigoni:2021guq}. The method developed in this paper allows for the systematic calculation of non-perturbative corrections as transseries, but their closed-form expressions are still missing.  Recently, in \cite{Bajnok:2022xgx,Bajnok:2024qro} the complete analytic transseries for densities of conserved charges in two-dimensional integrable quantum field theories in the presence of an external field have been explicitly constructed. It would be beneficial to improve our description and reach the same level of understanding in the present case.

Our results provide a new insight into the structure of non-perturbative contributions in strongly coupled four-dimensional supersymmetric gauge theories and offer promising avenues for further applications. In the AdS/CFT correspondence, the non-perturbative corrections are associated with world-sheet instantons. However, the number of examples where their contribution has been computed is very limited, and the existing technique can only be applied to obtain the first few terms in the strong coupling expansion. The method described in this paper allows to systematically compute the special class of observables in superconformal gauge theories for any coupling constant and to study their resurgence properties.  
 
\section*{Acknowledgements} 

We thank Benjamin Basso and Ivan Kostov for useful discussions and Annamaria Sinkovics for collaboration at an early stage. The research was supported by the Doctoral Excellence Fellowship Programme funded by the National Research Development and Innovation Fund of
the Ministry of Culture and Innovation and the Budapest University of Technology and Economics, under a grant agreement with the National Research, Development and Innovation Office (NKFIH). It was also supported by the research Grant No. K134946 of NKFIH and by the French National Agency for Research
grant “Observables” (ANR-24-CE31-7996).

\appendix

\section{Convergence of the weak coupling expansion}\label{app}

In this Appendix, we present a derivation of the relation \re{KL}. We can apply \re{eq:K_nm} and \re{K(x,y)} to express $\tr K^L(g)$ as $L-$fold convolution of the Bessel kernel
\begin{align}
\tr K^L(g) = \int_{0}^{\infty}dy_1\dots dy_L \, K(y_L,y_1)\chi\Big( {\sqrt{y_1}\over 2g}\Big)\dots K(y_{L-1},y_L)\chi\Big( {\sqrt{y_L}\over 2g}\Big)\,.
\end{align}
Changing the integration variables as $y_n=(2g x_n)^2$ (with $n=1,\dots,L$) and replacing $K(y_i,y_{i+1})$ with its representation \re{K(x,y)} we arrive at
\begin{align}\notag\label{trL}
\tr K^L(g) {}&=(4g^2)^L \int_0^\infty dx_1\dots dx_L \lr{ x_1\chi(x_1)\dots x_L\chi(x_L)}
\\
{}& \times \int_0^1 dt_1 \dots dt_L \lr{t_1 J_\ell(2g t_1 x_L)J_\ell(2g t_1 x_1) \dots t_L J_\ell(2g t_L x_{L-1})J_\ell(2g t_L x_L) }\,.
\end{align}
The analytic properties of $\tr K^L(g)$ are directly tied to the convergence of the integrals on the right-hand side. Let us examine the contribution to \re{trL} from large $x_i$. According to \re{chi-inf}, the product of the $\chi-$functions decreases exponentially fast whereas the product of the Bessel functions is rapidly oscillating in this region, rendering the integral well-defined for real $g$.  For pure imaginary values of the coupling, for $g=ih$ with $h$ real positive, the Bessel function 
has different behaviour at large $x$
\begin{align}\label{J-as}
J_\ell(2ih tx)=i^\ell I_\ell(2htx) = i^\ell {e^{2htx}\over 2\sqrt{\pi h t x}} + \dots\,,
\end{align}
where dots denote subleading terms.
Substituting this relation into \re{trL} and taking into account \re{chi-inf}, we find that for $t_i\to 1$ the integrand scales at large $x_i$ as $e^{(4h-1)x_i}$. 
Thus, the integral in \re{trL} is expected to develop a singularity at $h=1/4$, or equivalently $g^2=-1/16$.

To see this we substitute \re{J-as} and \re{chi-inf} into \re{trL} and obtain
\begin{align}\notag
\tr K^L(g) {}& = c^L (2ih)^{2L}\int_0^\infty dx_1 x_1 e^{-x_1} \dots dx_L  x_L e^{-x_L}
\\
{}& \times \lr{i^\ell\over 2\sqrt{\pi h}}^{2L} \int_0^1 dt_1 t_1 \dots dt_L t_L\, 
 \frac{e^{2 h t_1 x_1}}{\sqrt{t_1 x_1}}\frac{e^{2 h t_2 x_1}}{\sqrt{t_2 x_1}}\dots  \frac{e^{2 h t_L x_L}}{\sqrt{t_L x_L}}\frac{e^{2 h t_1 x_L}}{\sqrt{t_1 x_L}} +\dots
\end{align}
After integration over $x$'s we get
\begin{align}
\tr K^L(g) {}& = {c^L h^L\over \pi^L} (-1)^{L(1+\ell)} \kappa_L(h)+\dots\,,
\end{align}
where the notation was introduced for
\begin{align}\notag
\kappa_L(h) {}&= \int_0^1 {dt_1  \dots dt_L \over (1-2h(t_1+t_2)) \dots (1-2h(t_L+t_1)) } 
\\
{}&
 = -\frac{2^{L-1} \pi ^{L-\frac{3}{2}} \Gamma \left(\frac{L}{2}\right)}{\Gamma
   \left(\frac{L}{2}+\frac{1}{2}\right)} \log(1-4h) + O((1-4h)^0)\,,
\end{align}
where the second relation holds for $h\to 1/4$. Combining together the last two relations and replacing $h=-ig$, we arrive at \re{KL}.
 
\bibliography{bbk2}
\bibliographystyle{JHEP}  
 
\end{document}